%Paper: hep-th/9312172
%From: Jerome Gauntlett <jerome@yukawa.uchicago.edu>
%Date: Tue, 21 Dec 93 15:48:13 CST
%Date (revised): Tue, 21 Dec 93 16:26:32 CST
%Date (revised): Tue, 21 Dec 93 16:34:07 CST
%Date (revised): Mon, 27 Dec 93 17:11:09 CST

%-------------------------
% This paper uses harvmac
%-------------------------
%%%%%%%%%%%%%%%% READ THIS %%%%%%%%%%%%%%%% READ THIS %%%%%%%%%%%%%%%%%%%%%
%%%%%%%%%%%%%%%%%%%%%%%%%%%%%%%%%%%%%%%%%%%%%%%%%%%%%%%%%%%%%%%%%%%%%%%%%%%
% This paper has figures appended in  a second part as a uuencoded        %
% compressed tar file with instructions for unpacking.  They will be auto-%
% matically included in the text if you have a functioning epsf.tex.      %
% If you don't have that macro package (available from hep-th), or don't  %
% have the figure files, COMMENT OUT THE FOLLOWING LINE:                  %
\input epsf                                                               %
% If you do not already have epsf.tex (it comes with the dvips driver),   %
% you can print out the postscript files separately.                      %
%                                                                         %
% WARNING: there is more than one version of epsf.tex dated 18 Jul 1990 . %
% some figures will produce errors unless you have the most recent        %
% version. note that the version of epsf.tex that comes with the          %
% NeXTstep 2.0/2.1 distribution is not an up-to-date version. you should  %
% get epsf.tex from hep-th if your dvips is up-to-date but your epsf.tex  %
% is not.                                                                 %
%%%%%%%%%%%%%%%%%%%%%%%%%%%%%%%%%%%%%%%%%%%%%%%%%%%%%%%%%%%%%%%%%%%%%%%%%%%
\input harvmac
\noblackbox
\def\Title#1#2{\rightline{#1}\ifx\answ\bigans\nopagenumbers\pageno0\vskip1in
\else\pageno1\vskip.8in\fi \centerline{\titlefont #2}\vskip .5in}

%%%%%%%%%%%%%%%%%%
%
% Figure macros, SBG 5/93
%
\ifx\epsfbox\UnDeFiNeD\message{(NO epsf.tex, FIGURES WILL BE IGNORED)}
\def\figin#1{\vskip2in}% blank space instead
\else\message{(FIGURES WILL BE INCLUDED)}\def\figin#1{#1}%\epsfverbosetrue
\fi
\def\Fig#1{Fig.~\the\figno\xdef#1{Fig.~\the\figno}\global\advance\figno
 by1}
%
%  ifig   usage:
%
%         \ifig\figlabel{caption}{figfile}{vsize}
%
% where vsize is the desired vertical size of the figure in truein
%
\def\ifig#1#2#3#4{
\goodbreak\midinsert
\figin{\centerline{\epsfysize=#4truein\epsfbox{#3}}}
\narrower\narrower\noindent{\footnotefont
{\bf #1:}  #2\par}
\endinsert
}
%
%-------------------
% useful definitions
%-------------------
%
\def\p{\partial}
\def\d{\nabla}
\def\F{\bar F}
\def\G{\bar G}
\def\vp{\varphi}
\def\({\left (}
\def\){\right )}
\def\[{\left [}
\def\]{\right ]}
\def\cald{{\cal D}}
\def\calO{{\cal O}}
\def\RN{Reissner-Nordstr\"om}

\def\hf{{1\over2}}

\def\ajou#1&#2(#3){\ \sl#1\bf#2\rm(19#3)}

\def\phiT{{\widetilde \varphi}}
\def\hB{{\widehat B}}
\def\hq{{\widehat q}}
\def\frac#1#2{{#1 \over #2}}

\def\p{\partial}

\def\R{\hbox{\rm I \kern-5pt R}}

\def\l3{\lambda_3}

\hyphenation{par-am-et-rised}

%
%-------------------
%referencess
%-------------------
%
\lref\QEP{T.C. Bradbury,\ajou Ann. Phys. &19 (62) 323;
D.G. Boulware,\ajou Ann. Phys. & 124 (80) 169;
P. Candelas and D.W. Sciama,\ajou Phys. Rev. &D27 (83) 1715.}
\lref\brill{D. Brill and H. Pfister, \ajou Phys. Lett.  &B228 (89) 359;
D. Brill and G. Horowitz,\ajou Phys. Lett. &B262 (91) 437.}
\lref\BDDO{T. Banks, A. Dabholkar, M.R. Douglas, and M O'Loughlin,
\ajou Phys. Rev. &D45 (92) 3607.}
\lref\BaOl{T. Banks and M. O'Loughlin, \ajou Phys. Rev. &D47 (93) 540.}
\lref\DXBH{S.B. Giddings and A. Strominger,\ajou
Phys. Rev. &D46 (92) 627, hep-th/9202004.}
\lref\Hawktd{S.W. Hawking,\ajou Phys. Rev. Lett. &69 (92) 406.}
\lref\StTr{A. Strominger and S. Trivedi, ``Information consumption by
Reissner-Nordstrom black holes,'' ITP/Caltech preprint
NSF-ITP-93-15=CALT-68-1851, hep-th/9302080.}
\lref\SuTh{L. Susskind and L. Thorlacius,\ajou Nucl. Phys. &B382 (92) 123.}
\lref\dray{A. Ashtekar and T. Dray, \ajou Comm. Math. Phys. &79 (81) 581;
T. Dray, \ajou Gen. Rel. Grav. &14 (82) 109.}
\lref\dgt{H.F. Dowker, R. Gregory and J. Traschen, \ajou Phys. Rev. &D45 (92)
2762.}
\lref\HoWi{C.F.E. Holzhey and F. Wilczek, \ajou Nucl. Phys. &B380 (92) 447.}
\lref\unwald{W.G. Unruh and R.M. Wald,\ajou Phys. Rev. &D29 (84) 1047. }
\lref\rotbh{ A. Sen, \ajou Phys. Rev. Lett. &69, (92) 1006; J. Horne
and G. Horowitz \ajou  Phys. Rev.  &D46 (92) 1340.}
\lref\GGS{D. Garfinkle, S.B. Giddings and A. Strominger, ``Entropy in Black
Hole Pair Production'', Santa Barbara preprint UCSBTH-93-17,
gr-qc/9306023, to appear in Phys. Rev. D.}
\lref\kw{W. Kinnersley and M. Walker, \ajou Phys. Rev. &D2 (70) 1359.}
\lref\senny{S. Hassan and A. Sen, \ajou Nucl. Phys. &B375 (92) 103.}
\lref\he{J. Ehlers, in {\it Les Theories de la Gravitation},
(CNRS, Paris, 1959); B. Harrison, \ajou J. Math. Phys. &9 (68) 1744.}
\lref\CGHS{C.G. Callan, S.B. Giddings, J.A. Harvey and A. Strominger,
\ajou Phys. Rev. &D45 (92) R1005.}
\lref\ebh{S.B. Giddings and A. Strominger, \ajou Phys. Rev. &D46 (92) 627.}
\lref\kkbhs{H. Leutwyler, \ajou Arch. Sci. &13 (60) 549;
P. Dobiasch and D. Maison, \ajou Gen. Rel. and Grav. &14 (82) 231;
A. Chodos and S. Detweiler, \ajou Gen. Rel. and Grav. &14 (82) 879;
D. Pollard, \ajou J. Phys. A &16 (83) 565;
G.W. Gibbons and D.L. Wiltshire, \ajou Ann.Phys. &167
(86) 201;
erratum \ajou ibid. &176 (87) 393.}
\lref\gm{G.W. Gibbons and K. Maeda,
\ajou Nucl. Phys. &B298 (88) 741.}
\lref\ghs{D. Garfinkle, G. Horowitz, and A. Strominger,
\ajou Phys. Rev. &D43 (91) 3140, erratum\ajou Phys. Rev.
& D45 (92) 3888.}
\lref\AfMa{I.K. Affleck and N.S. Manton,
\ajou Nucl. Phys. &B194 (82) 38\semi
I.K. Affleck, O. Alvarez, and N.S. Manton,
\ajou Nucl. Phys. &B197 (82) 509.}
\lref\geroch{R.P. Geroch, \ajou J. Math. Phys. &8 (67) 782.}
\lref\raftop{R.D. Sorkin, in
{\sl Proceedings of the Third Canadian Conference on General
Relativity and Relativistic Astrophysics}, (Victoria, Can\-ada, May 1989),
eds. A. Coley, F. Cooperstock and B. Tupper (World Scientific, 1990).}
\lref\wheeler{J. Wheeler, \ajou Ann. Phys. &2 (57) 604.}
\lref\gwg{G.W. Gibbons,
in {\sl Fields and geometry}, proceedings of
22nd Karpacz Winter School of Theoretical Physics: Fields and
Geometry, Karpacz, Poland, Feb 17 - Mar 1, 1986, ed. A. Jadczyk (World
Scentific, 1986).}
\lref\CBHR{S.B. Giddings, ``Constraints on black hole remnants,'' UCSB preprint
UCSBTH-93-08, hep-th/9304027, to appear in {\sl Phys. Rev D}.}
\lref\garstrom{D. Garfinkle and A. Strominger,
\ajou Phys. Lett. &256B (91) 146.}
\lref\ernst{F. J. Ernst, \ajou J. Math. Phys. &17 (76) 515.}
\lref\rafKK{R. Sorkin, \ajou Phys. Rev. Lett. &51 (83) 87.}
\lref\grossperry{D. Gross and M.J. Perry, \ajou Nucl. Phys. &B226 (83) 29.}
\lref\hawk{S.W. Hawking in {\sl General relativity : an Einstein centenary
survey}, eds. S.W. Hawking, W. Israel (Cambridge University Press, Cambridge,
New York, 1979).}
\lref\schwinger{J. Schwinger,\ajou Phys. Rev. &82 (51) 664.}
\lref\melvin{M. A. Melvin, \ajou Phys. Lett. &8 (64) 65.}
\lref\Witt{E. Witten, \ajou Nucl. Phys. &B195 (82) 481.}
\lref\BOS{T. Banks, M. O'Loughlin and A. Strominger, \ajou Phys. Rev.
&D47 (93) 4476.}
\lref\ginsperry{P. Ginsparg and M.J. Perry, \ajou Nucl.Phys. &B222 (83) 245.}
\lref\Triv{S. Trivedi,\ajou Phys. Rev. &D47 (93) 4233.}
\lref\ChFu{S.M. Christensen and S. Fulling, \ajou Phys. Rev. &D15 (77) 2088.}
\lref\BGHS{B. Birnir, S.B. Giddings, J.A. Harvey and A. Strominger,
\ajou Phys. Rev. &D46 (92) 638.}
\lref\DGKT{H.F. Dowker, J.P. Gauntlett, D.A. Kastor and J. Traschen,
``Pair creation of dilaton black holes", to appear in Phys. Rev. {\bf D}.}
\lref\Cilar{S.B. Giddings, ``Comments on Information Loss and Remnants",
UCSBTH-93-35, hep-th/9310101.}
\lref\roternst{J.F. Plebanski and M. Demianski, \ajou Ann. Phys.
&98 (76) 98.}
\lref\bril{D. Brill, \ajou Phys. Rev. &D46 (92) 1560.}
\lref\kt{D.A. Kastor and J. Traschen,
``Particle Production and Positive Energy Theorems for Charged Black
Holes in De Sitter'',  UMHEP-399,  gr-qc/9311025, November 1992.}
%
%-------------------
% title page
%-------------------
%
\Title{\vbox{\baselineskip12pt
\hbox{EFI-93-74}
\hbox{UCSBTH-93-38}
\hbox{hep-th/9312172}}}
{\vbox{\centerline{ On Pair Creation of Extremal Black Holes }
\centerline{ and Kaluza-Klein Monopoles}
       }}
{
\baselineskip=12pt
\centerline{Fay Dowker,$^{1a}$
Jerome P. Gauntlett,$^2$ Steven B. Giddings,$^{1b}$
Gary T. Horowitz$^{1c}$}
\bigskip
\centerline{\sl $^1$Department of Physics}
\centerline{\sl University of California}
\centerline{\sl Santa Barbara, CA 93106}
\centerline{\it $^a$Internet: dowker@cosmic.physics.ucsb.edu}
\centerline{\it $^b$Internet: giddings@denali.physics.ucsb.edu}
\centerline{\it $^c$Internet: gary@cosmic.physics.ucsb.edu}
\bigskip
\centerline{\sl $^2$Enrico Fermi Institute, University of Chicago}
\centerline{\sl 5640 S. Ellis Avenue, Chicago, IL 60637 }
\centerline{\it Internet: jerome@yukawa.uchicago.edu}
\medskip
\centerline{\bf Abstract}
Classical solutions describing charged dilaton black holes
accelerating in a background magnetic field have recently
been found. They include
the Ernst metric of the Einstein-Maxwell
theory as a special case.
We study the extremal limit  of these solutions in detail,
both at the classical and quantum levels.  It is shown that near the
event horizon, the extremal solutions reduce
precisely to the static extremal black hole solutions.
For a particular value of the dilaton coupling, these
extremal black holes are five dimensional
Kaluza-Klein monopoles.
The euclidean
sections of these solutions can be interpreted as
instantons describing the pair creation of extremal black holes/Kaluza-Klein
monopoles in a magnetic field.
The action of these instantons is calculated and found to agree
with the Schwinger result in the weak field limit. For the euclidean Ernst
solution, the action for the extremal solution differs from that of the
previously discussed wormhole instanton by the Bekenstein-Hawking entropy.
However, in many cases quantum corrections become large in the vicinity of
the black hole, and
the precise description of the
creation process is unknown.
}

%\draftmode

%\Date{}

%
%---------------------------------------------
%

\newsec{Introduction}
The creation of particle-antiparticle pairs in a background field is
a common feature of quantum field theory. Schwinger \schwinger\
first studied this
process for electrons and positrons in a uniform electric field. This
was extended by Affleck and Manton \AfMa\
to the case of GUT monopole-antimonopole
production in a background magnetic field. In general relativity, the
analogue of a monopole is a magnetically charged black hole, and the question
naturally arises as to whether black holes can be pair produced by a
background magnetic field. Even though black holes and monopoles are
both ``solitons" in the sense of being static extended objects, there
are important
differences. First, the configuration of two black holes has a different
spatial topology than the vacuum. So  unlike the monopole case,
one cannot continuously deform one into
the other. Pair production of black holes is necessarily a topology
changing process.
 A second difference is simply that the fundamental quantum theory is
known for the case of monopoles (Yang-Mills-Higgs theory) but we
still lack a  quantum theory of gravity from which we can calculate
black hole pair creation  rates
 from first principles. Nevertheless, the previous
calculations were done in the context of an instanton approximation, and
it seems reasonable to hope that a similar approach will work for
black holes in a sum-over-histories framework for quantum gravity.

Affleck and Manton used an approximate instanton to estimate the pair
creation rate for GUT monopoles. As Gibbons first realized \gwg,
an exact instanton for the Einstein-Maxwell theory can be obtained
by analytically continuing a solution
found by Ernst almost twenty years ago \ernst. The Ernst solution describes two
oppositely charged black holes undergoing uniform acceleration in a background
magnetic field\foot{Ernst actually considered electric fields, but by duality,
that is equivalent to the magnetic fields we will consider here.}.
It describes the evolution of the black holes after their
creation. Regularity of the euclidean instanton turns out to
restrict the charge to mass
ratio of the black holes. Gibbons believed that only extremal black holes
could be created. But
Garfinkle and Strominger \garstrom\ found a regular instanton for which the
black holes were slightly non-extremal.
Furthermore, the horizons of the two black holes were identified
to form a wormhole in space.

For static charged black holes, the properties of the extremal solution
are quite different from the non-extremal one. In particular, the spatial
geometry of the extremal \RN\ metric resembles an infinite throat connected
onto an asymptotically flat region.  If extremal black holes of this
type can be pair created, one has the intriguing possibility
that an infinite volume of space could be quantum mechanically created in
a finite time. One of the aims of this paper is to investigate this
possibility. We explicitly  exhibit an extremal Ernst instanton
and study its properties.
We will show that as one approaches the horizon, the
solution reduces to the extremal \RN\ solution with its infinite
throat\foot{This can be compared to an instanton found by Brill \bril\
which describes the splitting of one throat into many. The present
instanton  differs in that it includes the asymptotic region and does not
require that a throat be present initially.}.
Furthermore, we will see that despite the infinite throats, the action
is finite and agrees with Schwinger's result for weak magnetic fields.
(This was also true for the non-extremal wormhole  \garstrom.)
However, higher order quantum corrections may become large down the throat and
modify the geometry significantly.

The situation in low energy string theory is similar. Extremal magnetically
charged black holes have a spatial geometry which is identical to that of
the \RN\ solution \refs{\gm,\ghs}. An
effort was made to calculate the production rate for
black holes in string theory \BOS\ by constructing an
instanton in a hybrid two- and four-dimensional theory which was conjectured
to approximate the full theory.
The resulting configuration described non-extremal black holes
with their horizons identified to form a wormhole. Arguments
have been made that  instantons corresponding to the pair creation of
extremal black holes  do not  exist.  We will
see that this is
incorrect; indeed, in the string case, the natural analogue
of the wormhole instanton develops an infinite throat and becomes extremal.
As in the \RN\ case, this reduces to the static solution
far down the throat.

A convenient way to treat both the Einstein-Maxwell and low energy  string
theories simultaneously is to consider the action
\eqn\action{
S={1\over 16 \pi}\int d^4x
{\sqrt {-g}}\left[R-2(\nabla\phi)^2-e^{-2a\phi}F^2\right].
}
 For $a=0$ this is the standard Einstein-Maxwell theory coupled to a massless
scalar field $\phi$. By the no-hair theorems, $\phi$ must be constant for
solutions describing static black holes. For $a=1$, $S$ is part of the action
describing the low energy dynamics of string theory. In this case,
$\phi$ is not constant outside a charged black hole. For some physical
questions it is more appropriate to use the conformally rescaled metric
$\tilde g_{\mu\nu} = e^{2\phi} g_{\mu\nu}$ which is called the string metric.
It is with respect to this metric that the extremal black holes have infinite
throats.
The value $a=\sqrt{3}$ is also of special interest since this corresponds to
standard
Kaluza-Klein theory, i.e. \action\ is equivalent to the
five-dimensional vacuum Einstein action for geometries with a spacelike
symmetry.

The equations of motion which follow from this action are
\eqn\eom{\eqalign{
  & \d_\mu (e^{-2a\phi}F^{\mu\nu}) = 0  \cr
   &\d^2 {\phi}+ {a\over 2} e^{-2a\phi}F^2 = 0  \cr
    &R_{\mu\nu} = 2 \d_\mu \phi  \d_\nu \phi  +2 e^{-2a\phi} F_{\mu\rho}
      {F_\nu}^\rho  -\half g_{\mu\nu} e^{-2a\phi} F^2.  \cr} }
Recently a solution to these equations was found (for all values of $a$) which
generalizes the Ernst solution \DGKT.
These dilaton Ernst solutions describe
two oppositely charged black holes
undergoing uniform acceleration in a background magnetic field.
 They contain a boost-like symmetry which allows one to analytically
continue to obtain euclidean instantons.
Regular instantons describing the pair creation of non-extremal
black holes with their horizons identified were constructed in
\DGKT\
for $0\le a< 1$ and shown not to exist for $a\ge1$.
In this paper we will choose the parameters so that
the black holes are extremal and study the resulting instantons for all
values of $a$.
As anticipated by Gibbons \gwg,
for $a=0$ the instanton is completely regular.
For $a=1$, the string metric corresponding to the extremal instanton
is  also regular, and
for $a=\sqrt 3$, the corresponding
five dimensional metric is again regular.

This paper is organized as follows. We begin, in Section 2, by
describing the general features of the lorentzian dilaton Ernst solution
and investigate its extremal limit.
In this limit, it is shown that as one approaches
the black hole, the solution reduces exactly to the static
black hole solution.
Thus for $a=0$ and $a=1$, the classical
extremal accelerated black holes have
infinite throats, just as in the static case.
In fact, we will
argue that there is a sense in which the extremal black hole is
{\it{not}}
accelerating despite the presense of the magnetic field. However, the
region around the black hole is accelerating.

In Section 3 we study
the corresponding euclidean instanton. We will show that this describes
creation of a pair of extremal black holes for each value of $a$.
The throats are not identified
to form a wormhole.
For fixed values
of the physical charge, $\widehat q$, and magnetic field $\widehat B$,
we compute the exact action for both the extremal
and wormhole instantons.
For weak fields, i.e. to leading order in $\widehat q\hB$, the
action reduces to the action found by Schwinger, for all values of $a$
and for both types of instantons. To the next order in
$\widehat q\hB$ we find that, for  $a=0$, the action of the wormhole
instanton is less than that of the extremal instanton
by the Bekenstein-Hawking
entropy $A/4$ of the extremal black holes.
We do not understand the physical significance of this
intriguing result which is reminiscent of \GGS\ in which it was found
that the action of the wormhole instanton is less than that of an
instanton describing pair creation of GUT monopoles by the same
entropy term.  This does suggest suppression of extremal pair creation relative
to that of wormholes, but quantum corrections can contribute at the same
order in $\widehat q\hB$.
For $ a \ne 0$ we find that the actions of the extremal and
wormhole instantons agree  even to next order in $\widehat q\hB$.
This is consistent with the $a=0$ result since in this case,
the area of the horizon
shrinks to zero size in the extremal limit. (The area of the horizon in
the nearly extremal wormhole instanton is non-zero, but higher order.)

Section 4 contains a discussion of the effect of quantum fluctuations
about the classical solution. This question is of interest in both the
lorentzian and euclidean contexts. Since Hawking's discovery of black hole
evaporation, there has been extensive discussion of quantum fields around
static black holes, and of observers accelerated through the vacuum. When the
black holes themselves are accelerating, these two subjects are combined
in a fundamental way. By using results from these two areas, we will argue
that the back-reaction may become strong
down the infinite throats. For the euclidean
instanton, the quantum fluctuations
appear to affect the infinite throats as well, and may substantially alter
the production rate.

The extremal  dilaton Ernst
solution is of special interest  for $a=\sqrt 3$, the
Kaluza-Klein case.
This
is studied in detail in Section 5.
It is known that the extremal magnetically charged black hole for
$a=\sqrt 3$
is the Sorkin-Gross-Perry monopole \refs{\rafKK,\grossperry}. So
the lorentzian solution describes two (oppositely charged)
monopoles accelerating in
a uniform magnetic field. The euclidean instanton describes pair creation of
Kaluza-Klein monopoles.
Even though the four dimensional metric describing an extremal magnetically
charged black hole
is singular, the corresponding five dimensional metric turns out to
be completely regular. So the five dimensional instanton is also regular.
Unlike
the previous cases, the quantum corrections should remain small everywhere,
so the action for the instanton should give a good approximation to the
pair creation rate.  We conclude in Section 6 with a discussion of
some open problems.

Before proceeding to the accelerated black hole solutions, we briefly review
the static dilaton black hole and background magnetic field solutions
to \eom.
The black hole is given by \refs{\gm,\ghs}
\eqn\dbhs{
\eqalign{
& ds^2=-\lambda^2dt^2+\lambda^{-2}dr^2+R^2(d\theta^2 +
\sin^2\theta d\varphi^2)
 \cr
 & e^{-2a\phi}=\left(1-{r_-\over r}\right)^{2a^2\over(1+a^2)},\qquad
 \qquad A_\varphi=q(1 - {\rm cos}\theta)\cr
 &\lambda^2=\left(1-{r_+\over r}\right)
 \left(1-{r_-\over r}\right)^{{(1-a^2)\over (1+a^2)}},\qquad
 R^2=r^2\left(1-{r_-\over r}\right)^{2a^2\over (1+a^2)}.\cr}
 }
If $r_+ > r_-$, the surface $r=r_+$ is the event horizon.
 For $a=0$, the surface $r=r_-$ is the
 inner Cauchy horizon, however for $a>0$
 this  surface is singular.
 The parameters $r_+$ and $r_-$ are related to the ADM mass $m$
 and total charge $q$ by
\eqn\mass{
m={r_+\over 2} + \left ({1-a^2\over 1+a^2}\right ){r_-\over 2},\qquad
q=\left({r_+r_-\over 1+a^2}\right)^{1\over 2}.}
The extremal limit occurs when $r_+=r_-$.
As one approaches this extremal limit,
the Hawking temperature goes to zero when $a<1$, approaches a non-zero constant
when $a=1$ and
diverges when $a>1$.

The solution
describing the background magnetic field was found by Gibbons and Maeda \gm\
and is given by\foot{The gauge
potential given here differs from that in \DGKT\ by a gauge transformation
so that $A_\mu$ is regular on the axis $\rho=0$.}
\eqn\dmelv{
\eqalign{
&ds^2=\Lambda^{2\over 1+a^2}\left[-dt^2+dz^2+d\rho^2\right]
+\Lambda^{-{2\over 1+a^2}}\rho^2d\varphi^2\cr
&e^{-2a\phi}=\Lambda^{2a^2\over 1+a^2},\qquad
A_\varphi={B\rho^2\over 2\Lambda}\cr
&\qquad \Lambda=1+{(1+a^2)\over 4}B^2\rho^2\ .\cr}
}
It is a generalization of Melvin's magnetic
universe \melvin. The square of the Maxwell field is $F^2 = 2B^2/\Lambda^4$,
which is a maximum
on the axis $\rho=0$ and decreases  to zero at infinity. The parameter
$B$ labels the strength of the magnetic field.
For $a > 0 $, the dilaton is zero on the axis but diverges
to minus infinity as $\rho \rightarrow \infty$.

\newsec{Dilaton Ernst metrics}
\subsec{General properties}
The dilaton Ernst solutions to \eom\ constructed in \DGKT\
represent two oppositely charged dilaton black
holes uniformly accelerating in a background magnetic field.
They generalize the Einstein-Maxwell ($a=0$) solutions found by Ernst \ernst.
They are:
\eqn\dernst{
\eqalign{
&ds^2=(x-y)^{-2}A^{-2}\Lambda^{2\over 1+a^2}
\left[F(x)\left\{G(y)dt^2-G^{-1}(y)dy^2\right\}
+F(y)G^{-1}(x)dx^2\right]\cr &\qquad +
(x-y)^{-2}A^{-2}\Lambda^{-{2\over 1+a^2}}F(y)G(x) d\varphi^2\cr
&e^{-2a\phi}=e^{-2a\phi_0}\Lambda^{2a^2\over 1+a^2}
{F(y)\over F(x)},\quad
A_\varphi=-{2e^{a\phi_0}\over (1+a^2)B\Lambda}\[1+{(1+a^2)\over 2}Bqx\]
+k\cr
}
}
where the functions $\Lambda \equiv\Lambda(x,y)$,
 $F(\xi)$ and $G(\xi)$ are given by
\eqn\fns{\eqalign{
&\Lambda=\[1+{(1+a^2)\over 2}Bqx\]^2+{(1+a^2)B^2\over 4A^2(x-y)^2}
G(x)F(x) \cr
&F(\xi)=(1+r_-A\xi)^{2a^2\over (1+a^2)}\ \cr
&G(\xi)=(1-\xi^2-r_+A\xi^3)(1+r_-A\xi)^{(1-a^2)\over
(1+a^2)}\ .\cr
}}
and $q$ is related to $r_+$ and $r_-$ by \mass.
The constant $k$ in the expression for the gauge field is introduced
so that the Dirac string of the magnetic field of a monopole
is confined to one axis. The constant $\phi_0$ in the solution for the
dilaton determines the value of the dilaton at infinity. Although one could
keep this as a free parameter, we will fix it so that the dilaton vanishes
on the axis at infinity in agreement with \dmelv. The values of both
$k$ and $\phi_0$ will be given below.

The solution \dernst\ depends
on four other parameters, $r_\pm,A,B$. Defining $m$ and $q$ via \mass\
we can loosely
think of these parameters together with $A,B$
as denoting the mass, charge and acceleration
of the black
holes and the strength of the magnetic field which is accelerating them,
respectively.
We emphasize, however, that this is heuristic since, for
example, the mass and acceleration are not in general
precisely defined and,
further, we will see that $q$ and $B$ only approximate the physical
charge $\widehat q$ and magnetic field $\widehat B$ in the limit
$r_\pm A\ll1$.

It is convenient to introduce the following notation.
Define $\xi_1\equiv -{1\over r_-A}$ and let $\xi_2\le\xi_3<\xi_4$
be the three roots  of the cubic in $G$.
The functions $F(\xi)$, $G(\xi)$ then take the form
\eqn\fandg{\eqalign{
&F(\xi)=(r_-A)^{2a^2\over1+a^2}(\xi-\xi_1)^{2a^2\over 1+a^2}\cr
&G(\xi)=-(r_+A)(r_-A)^{(1-a^2)\over(1+a^2)}
(\xi-\xi_2)(\xi-\xi_3)(\xi-\xi_4)(\xi-\xi_1)^
{(1-a^2)\over (1+a^2)}.\cr
}}
We restrict the range of the parameters $r_+$ and $A$ so that
$r_+A\le2/(3\sqrt{3})$,
so that the $\xi_i$ are all real; the limit $r_+A=2/(3\sqrt{3})$
corresponds to $\xi_2=\xi_3$.
We also restrict the parameter $r_-$
so that $\xi_1 \le \xi_2$.

The metric \dernst\ has two Killing vectors,
${\partial\over\partial t}$ and ${\partial\over\partial\varphi }$.
The surface $y=\xi_1$ is
singular for $a>0$, as can be seen from the square of the field strength.
This surface is analogous to the singular surface
(the ``would be'' inner horizon) of the dilaton black holes.
The surface $y=\xi_2$ is the black hole horizon and the surface
$y=\xi_3$ is the acceleration horizon; they are both Killing horizons for
${\partial\over\partial t}$.

The coordinates $(x,\varphi)$ in \dernst\ are angular coordinates.
To keep the signature of the metric fixed, the coordinate $x$
is restricted to the range $\xi_3 \le x\le\xi_4$ in which $G(x)$ is positive.
Due to the conformal factor $(x-y)^{-2}$ in the metric,
spatial infinity is reached by fixing $t$ and
letting both $y$ and $x$ approach
$\xi_3$.  Letting $y\rightarrow x$ for $x\ne\xi_3$ gives null or
timelike infinity \dray .
Since $y\rightarrow x$ is infinity, the range of the coordinate $y$ is
$-\infty<y<x$ for $a=0$, $\xi_1<y<x$ for $a>0$.

The norm of the
Killing vector ${\partial\over\partial\varphi }$ vanishes at $x=\xi_3$ and
$x=\xi_4$, which correspond to the poles of the spheres surrounding the black
holes.   The axis $x=\xi_3$ points along the symmetry axis
towards spatial infinity.  The axis $x=\xi_4$ points towards the other black
hole. The coordinates we are using only cover one region of
spacetime containing one of the black holes.
The Dirac string singularities attached to the monopoles will be taken to
lie along the axis $x=\xi_4$; this is accomplished by fixing the constant
$k$ so that $A_\varphi(x=\xi_3)=0$.

As discussed in \DGKT,
to ensure that the metric is free of conical singularities at both poles,
$x=\xi_3, \xi_4$,
we must impose the condition
\eqn\nonodes{G^\prime(\xi_3)\Lambda(\xi_4)^{2\over 1+a^2}
= -
G^\prime(\xi_4)\Lambda(\xi_3)^{2\over 1+a^2}.}
where\foot{It follows from \fns\ that
when $x$ is equal to
a root of $G(x)$, $\Lambda(x,y)$ is independent of $y$.  So $\Lambda(\xi_i)$
are constants.} $\Lambda(\xi_i)\equiv \Lambda(x=\xi_i)$.
It  will be convenient to define
\eqn\defl{L \equiv \Lambda^{1\over 1+a^2}(\xi_3)\ .}
When \nonodes\ is satisfied, the spheres are regular as long as $\vp$
has period
\eqn\phiperiod{\Delta\vp={4\pi L^2\over
G^\prime(\xi_3) }\ .}
The condition \nonodes\
can be readily understood in the limit $r_+ A\ll 1$, which
implies $r_- A\ll 1$. In
this case one has the expansions
\eqn\xiexp{\eqalign{\xi_2 &= -{1\over r_+ A} + r_+ A + \cdots\cr
\xi_3&= -1 -{r_+ A\over 2} + \cdots\cr
\xi_4&= 1-{r_+A\over 2} +\cdots .}}
 Substituting the expressions
\fns\ and  \fandg\ into \nonodes\ and
expanding to leading order in $r_+ A$ gives Newton's law,
\eqn\Newt{mA\approx qB\ ,}
where we have used \mass\ to replace $r_\pm$ with $m,q$. This is true for
all $a$. More generally,
the condition \nonodes\ reduces the number of free
parameters for the solution to three  by relating the
acceleration to the magnetic field, mass, and charge.

The appearance of Newton's law strongly suggests a regime in which the
solution closely approximates a point particle
moving in the Melvin background.  Indeed, the point particle limit is given
by $r_+ A\ll1$, since this corresponds to a black hole small on the scale
set by the magnetic field.
In this limit, and taking $|r_+Ay|\ll1$, one finds that
$G(\xi) \approx 1 - \xi^2$,
$F(\xi) \approx 1$ and the solution
\dernst\ reduces to the form
\eqn\Ernslim{\eqalign{ds^2 \approx &{\Lambda^{2\over 1+a^2}
\over A^2 (x-y)^2} \left[ (1-y^2)
 dt^2 - {dy^2\over  (1-y^2)} + {dx^2\over
(1-x^2) }\right]\cr & + \Lambda^{-{2\over 1+a^2}}{1-x^2 \over  (x-y)^2 A^2}
d\vp^2}}
with
\eqn\Llim{\Lambda\approx 1+ {(1+a^2)B^2\over 4} {1-x^2\over A^2
(x-y)^2}\ .}
The coordinate transformation
\eqn\coordlim{\rho^2 = {1-x^2 \over  (x-y)^2 A^2}\ ,\ \zeta^2 = {y^2-1\over
(x-y)^2 A^2}}
simplifies this to
\eqn\Melvl{ds^2 \approx \Lambda^{2\over 1+a^2}\left[ -\zeta^2 dt^2
+d\zeta^2
+ d\rho^2
 \right] + \Lambda^{-{2\over 1+a^2}} \rho^2 d\vp^2 \ }
with $\Lambda$ given in \dmelv.  The dilaton and gauge fields are likewise
found to be
\eqn\digalim{A_\vp\approx e^{a\phi_0} {B\rho^2\over 2\Lambda}\ ,\
e^{-2a\phi} \approx e^{-2a\phi_0} \Lambda^{2a^2\over 1+a^2}\ ,}
where $k$ has been chosen so that $A$ is regular on the axis $\rho=0$.

Eqs.~\Melvl,\digalim\ give the dilaton Melvin solution \dmelv, expressed in
Rindler
coordinates, up to the arbitrary shift of the asymptotic value of the
dilaton.  The standard form follows using
the
coordinate transformation $\widehat t=\zeta {\rm sinh}t,\,
z=\zeta{\rm cosh}t$.
Thus the coordinate $t$ in the dilaton Ernst solution is the
analogue of Rindler time.  The subleading terms in \Melvl\ become important
when $y\approx -1/r_+ A$, which, for small
$r_+ A$  corresponds to $\rho\approx 0$,
$\zeta\approx 1/A$ --
the trajectory of the black hole.  The asymptotic limit \Melvl\ is obtained
a distance of order $r_+$
 from the black hole, as expected.  These features are
illustrated in fig.~1.

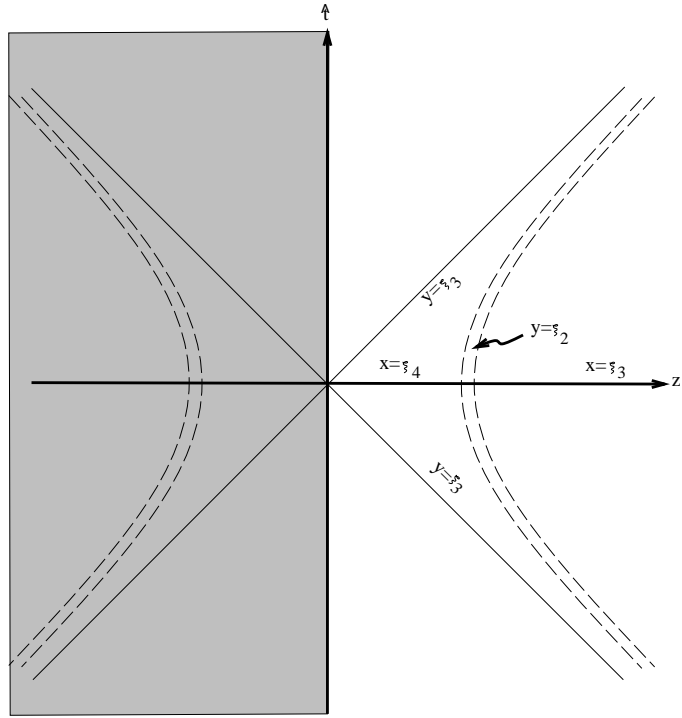
\ifig{\Fig\Ppart}{The $z,{\hat t}$ plane of the dilaton Ernst
solution, in the limit $r_+A\ll1$.
The dotted lines indicate a region of size $\sim r_+$
surrounding the black hole at $y\sim \xi_2$;
inside this region the geometry is
approximately that of the black hole.  The black hole moves on a trajectory
with acceleration $A$, and the acceleration horizon is given by $y=\xi_3$.
The coordinates used in \dernst\ cover only the unshaded part
of the figure.}{pcebh.fig1}{3.8}

The relation to Melvin is not restricted to the point-particle
limit; even away from this limit \dernst\ becomes Melvin
at large spacelike distances.  This corresponds to
$x,y \rightarrow \xi_3$. One way to show that the metric \dernst\
approaches \dmelv\ asymptotically was given in \DGKT. A somewhat simpler
approach
is to change coordinates from $(x,y,t,\vp)$ to $(\rho,\zeta,\eta,\phiT)$ using
\eqn\coordch{\eqalign{&x-\xi_3 = {4F(\xi_3) L^2
\over G'(\xi_3) A^2} {\rho^2\over (\rho^2+\zeta^2)^2}\quad ,\quad
\xi_3-y
={4F(\xi_3) L^2
\over G'(\xi_3) A^2} {\zeta^2\over (\rho^2+\zeta^2)^2} \cr
&t= {2\eta\over  G'(\xi_3)}\quad ,\quad \vp = {2 L^2
\over G'(\xi_3) }\phiT\ }}
Note that  $\eta, \phiT$  are related to $t,\vp$ by a simple rescaling and
that $\phiT$ has period $2\pi$ due to \phiperiod.
For large $\rho^2+\zeta^2$, the dilaton Ernst metric reduces to
\eqn\blip{
ds^2  \to \tilde\Lambda^{2\over 1+a^2}\left(-\zeta^2d
\eta^2+d\zeta^2+d\rho^2\right)+\tilde\Lambda^
{-{2\over 1+a^2}}\rho^2 d\widetilde\varphi^2}
where
\eqn\blob{
\eqalign{&\tilde\Lambda=\left(1+{1+a^2\over 4}\widehat B^2\rho^2\right)\
,\cr
&\widehat B^2=
{B^2G^{\prime}(\xi_3)^2\over 4L^{3+a^2}}\ .}
}
Again we recover the dilaton Melvin metric in Rindler coordinates.

The asymptotic form of the dilaton and gauge potential are
\eqn\dilga{ \eqalign{&e^{-2a\phi}\to
L^{2a^2} \tilde\Lambda^{2a^2\over 1+a^2} e^{-2a\phi_0} \cr
&A_{\tilde\varphi}\to L^{-a^2} e^{a\phi_0}
{{\widehat B} \rho^2 \over  2 {\tilde \Lambda}}\
.}}
This is equivalent to the standard background magnetic field solution \dmelv\
provided we choose
\eqn\dilzero{e^{a\phi_0} = L^{a^2}}
 We will take the constant $\phi_0$ to be fixed
at this value in the remainder of the paper. We can now see that the physical
magnetic field is $\widehat B$ given by \blob.  Using \xiexp\ we note
that in the limit $r_\pm A\ll 1$,
$\widehat B\approx B$.

The physical charge of the black hole is defined by $\hq = {1\over 4 \pi}
\int F$ where the integral is over any two sphere surrounding the black hole.
For the dilaton Ernst solution \dernst, one obtains
\eqn\pchge
{\widehat q = q
{ L^{3+a^2\over 2} (\xi_4-\xi_3) \over G'(\xi_3) (1+ {1+a^2 \over 2} qB \xi_4)}
\ .}
In the weak field limit $r_\pm A\ll 1$, $\hq \approx q$.
Using \blob, the product of the physical charge and magnetic field is
\eqn\hatbq{ \hq \hB = {qB (\xi_4-\xi_3) \over 2(1+ {1+a^2 \over 2} qB \xi_4)}
\ .}
This will be useful shortly.

\subsec{The limit $\xi_1=\xi_2$: accelerating extremal black holes}

Since $y=\xi_2$ is the event horizon and $y=\xi_1$ is an inner horizon
($a=0$) or singularity ($a >0$), it follows that the extremal limit
of the dilaton Ernst solutions is given by choosing the parameter $r_-$
 so that $\xi_1=\xi_2$. Recalling the regularity condition \nonodes,
it follows that the extremal solutions are described by two
parameters which we can take to be the physical charge and
magnetic field. In this section we will show that as $y\rightarrow \xi_2$
the extremal solutions become spherically symmetric, and approach the
static black hole  solutions \dbhs\ with $r_-=r_+$. This surprising
result has
a number of consequences which we will discuss.

Since the derivation involves considerable algebra, we will simply indicate
the main steps involved. The first step is to show that
with $\xi_1 =\xi_2$, one can divide the condition for  no nodal
singularities \nonodes\ by $\xi_4- \xi_3$
to obtain
\eqn\cons{
1+(1+a^2)Bq\xi_2+{1\over 4}(1+a^2)^2B^2q^2(\xi_2\xi_3+\xi_2\xi_4-
\xi_3\xi_4) = 0  \ .
}
Taking the limit $y\rightarrow \xi_2$ and using this equation,
the function $\Lambda$ in \dernst\
becomes
\eqn\al{
\Lambda\to\alpha(x-\xi_2)
}
where
\eqn\alp{
\alpha=(1+a^2)Bq + \[{(1+a^2)Bq\over 2}\]^2(\xi_3+\xi_4)
\ .}
One can then show that the dilaton Ernst metric \dernst\ tends to
\eqn\down{
\eqalign{
&ds^2 \rightarrow ds_0^2\cr
& =
-(r_+r_-) \alpha^{2\over 1+a^2} (\xi_4-\xi_2) (\xi_3-\xi_2)
(y-\xi_2)^{2\over 1+a^2} dt^2  \cr
&
+{ \alpha^{2\over 1+a^2} r_-^{3a^2-1\over
1+a^2}
\over A^{4\over 1+a^2} r_+}\Biggl[ {dy^2\over
(\xi_4-\xi_2) (\xi_3-\xi_2)
(y-\xi_2)^{2\over 1+a^2}}
\cr&
+ {(y-\xi_2)^{2a^2\over 1+a^2}
\over (x-\xi_2)^2}\left({dx^2\over (x-\xi_3)(\xi_4-x)} + {4
(x-\xi_3)(\xi_4-x) \over (\xi_4-\xi_3)^2 } d\phiT^2\right)\Biggr]\ }}
where we have used the coordinate $\phiT$ introduced in \coordch.

At this stage it is not obvious that the $x,\phiT$ part of the
metric \down\ corresponds
to the round two sphere, but the coordinate change
\eqn\xcoordc{ x={1\over 2} \left[ \xi_3+\xi_4 -
{\xi_4-\xi_3
+(\xi_4+\xi_3-2\xi_2)\cos\theta\over \cos\theta +{\xi_4+\xi_3-2\xi_2\over
\xi_4-\xi_3}}\right]}
puts it in the standard form, with polar coordinates $\theta,\phiT$.
The final step consists of introducing a new variable
\eqn\rpdef{
{\widehat r}_+^2 = {\alpha^{2\over 1+a^2} r_-^{3a^2-1\over 1+a^2}
(-\xi_2)^{2a^2 \over 1+a^2}\over A^{4\over 1+a^2} r_+ (\xi_4-\xi_2)
(\xi_3-\xi_2) }\ , }
and new coordinates
\eqn\coordcx{\eqalign{t'&=\sqrt{r_+ r_- (\xi_4-\xi_2) (\xi_3-\xi_2) }
(-\alpha\xi_2)^{1\over 1+a^2} t\cr
y&={{\widehat r}_+\over r}\xi_2\ .}}
In the limit $y\rightarrow\xi_2$, these coordinates simplify \down\ to
\eqn\throata{ds_0^2 =
 -\left(1-{{\widehat r_+}\over r}\right)^{2\over{1+a^2}} {dt'}^2 +
\left(1-{{\widehat r_+}\over r}\right)^{-2\over{1+a^2}}dr^2 +
{\widehat r_+}^{2}\left(1-{{\widehat r_+}\over r}\right)^{{2a^2}\over{1+a^2}}
d\Omega^2\ .}
The behavior of the
other fields can also be worked out in the limit $y\rightarrow\xi_2$.
One obtains:
\eqn\throatb{\eqalign{&
A_{\widetilde\varphi} \rightarrow  \widehat q\left(1- {\rm cos}\theta
\right)\ ,\cr
&e^{-2a\phi} \rightarrow  e^{-2a\phi_0}
\left(-\alpha \xi_2\right)^{{2a^2}\over{1+a^2}} \left( 1 -{{\widehat r_+}
\over r}\right)
^{{2a^2}\over{1 +a^2}}\ , } }
where
\eqn\relations{
\widehat q = {{\widehat r_+}\over{\sqrt{1+a^2}}} e^{a\phi_0}
\left(-\alpha \xi_2\right)^{{-a^2}\over{1+a^2}}
}
This agrees with the  extremal static dilaton black hole (given by \dbhs\
with $r_+ = r_-$), in the limit $r \rightarrow r_+$.\foot{Equations \throata\
and \throatb\ are exactly the extremal static solution except for a constant
shift in the dilaton. }
 Using \al\ with $x=\xi_3,\xi_4$, \fandg\ and \mass\ one can show
that
$\widehat q$   agrees with \pchge\ when the black hole is
extremal $\xi_1 = \xi_2$.

The fact that the extremal dilaton Ernst solution approaches the static
black hole as $y \rightarrow \xi_2$ has several consequences. {}First, all
the geometric properties of the extremal static solutions near the horizon
carry over immediately to the accelerated case. In particular,
for $a=0$, a constant-$t$ slice of the solution has
an infinitely long throat. {}For $a=1$, the string metric
$d\tilde s^2 = e^{2\phi} ds^2$ also has an infinite throat in which
the solution takes the form of the linear dilaton
vacuum. {}For $a=\sqrt{3}$, the four-metric, dilaton and gauge field
together make up the five dimensional metric of the Kaluza-Klein
monopole (see Section 5).

A second consequence is that there is a sense in which the extremal
black holes are not accelerating. For $a=0$, this is suggested by the
fact that the event horizon is exactly spherical. But a more convincing
argument comes from examining the acceleration of a family of observers
near the horizon whose four velocities are proportional to $\p /\p t$.
{}For the static black hole, the acceleration of these observers approaches
the finite limit $1/q$ as they approach the horizon. (This is
related to the fact that the surface gravity vanishes for extremal
black holes and is in contrast to the non-extremal case in
which the acceleration diverges.)
If one computes the acceleration of these observers for the Ernst solution,
one again finds that it approaches $1/\widehat q$ as $y \rightarrow \xi_2$
independent of direction. Although this particular argument cannot be
extended to $a >0$ since the acceleration (in the Einstein metric)
now diverges for the static
extremal solution
near the horizon, other arguments can be made. For example,
when $a=1$, in the string metric the acceleration of these observers
tends to zero down the throat. In addition,
when $a=\sqrt{3}$, $y=\xi_2$ is a regular
origin in the five dimensional Kaluza-Klein solution, and one can show that
its worldline is a geodesic!

Even though the black hole itself is not accelerating, the region around
the black hole is. This is clear from the relation between the  solution
and the dilaton Melvin solution in accelerating coordinates discussed in
the previous subsection. In terms of the infinite throats, one might say
that the mouth of the throat is accelerating  while the region down the
throat is not.

A final comment concerns the physical charge and magnetic field.  Consider
the product $\widehat q \widehat B$. This is small in the weak field limit,
which corresponds to $\xi_2$ being large and negative. What happens
when, instead, $\xi_2$ approaches $\xi_3$? The two roots
$\xi_2$ and $\xi_3$ can approach each other only if $\xi_2,\xi_3 \to -\sqrt 3$
and $\xi_4 \to \sqrt 3/2$. Using this,  eq. \hatbq, and the no strut condition
\cons, one can show
\eqn\bqlim{\widehat q \widehat B \rightarrow {1\over 1+a^2}\ .}
Thus there is an upper bound on the product of the charge and magnetic field.
Roughly speaking, since the size of an extremal black
hole is $\sim\widehat q$ and the width
of the Melvin flux tube is $\sim 1/(\widehat B {\sqrt {1+a^2}})$, one of the
consequences of \bqlim\ is that the black holes are moving in
flux tubes wider than themselves.
The limit  $\xi_2 \rightarrow \xi_3$ corresponds to the event horizon
approaching the acceleration horizon. Since we have assumed the
black holes are extremal, $\xi_1 = \xi_2$,
this corresponds to a ``triple point" where three roots coincide.  If one
relaxes the condition $\xi_1 = \xi_2$ it appears that the bound on
$\widehat q \widehat B$ is even lower, which is consistent with the fact that
the event horizon is larger than the charge and hits the acceleration
horizon at a smaller value of $\widehat q$.
What happens if one takes a charged black hole and turns up the magnetic
field larger than the bound \bqlim? It would appear that this situation
is no longer
described by the class of solutions \dernst. The question of what
happens physically is currently under investigation.

\newsec{Dilaton Ernst Instantons}
Euclideanizing \dernst\ by setting $\tau=it$, we find that
another condition must be imposed
on the parameters
in order to obtain a regular solution. Two distinct ways that
this may be achieved were discussed in \DGKT\ and are reviewed
in the first subsection below. These include the wormhole instantons.
There is a third
 way which leads to  the extremal instantons
and is described in the second subsection. The calculation of the
action for the wormhole and extremal instantons is given in the
third subsection.

\subsec{Wormhole Instantons}

In the lorentzian solutions, the vector $\p/\p t$ is timelike only for
$\xi_2 <y < \xi_3$.
In \DGKT\  the restriction $\xi_1<\xi_2$
was made so that the
Einstein metric had a regular horizon for $a>0$.
In this case,
one must impose a condition on the parameters in
order to
eliminate conical singularities in the euclidean solution
at both the black hole ($y=\xi_2$) and
acceleration ($y=\xi_3$) horizons
with a single choice of the period of $\tau$.
This is equivalent to demanding that the
Hawking temperature of the black hole horizon equal the Unruh
temperature of the acceleration horizon.

In terms of the metric function $G(y)$ appearing in \dernst, the period of
$\tau$ is taken to be
\eqn\tauper{\Delta \tau ={4\pi\over G^\prime(\xi_3)}}
and the constraint
is
\eqn\regular{
G^\prime(\xi_2) =- G^\prime(\xi_3), }
yielding
\eqn\roots{
\left({\xi_2-\xi_1\over \xi_3-\xi_1}\right)^{1-a^2\over 1+a^2}(\xi_4-\xi_2)
(\xi_3 -\xi_2)
=(\xi_4 - \xi_3)(\xi_3 -\xi_2). }
With $\xi_1<\xi_2$ there are two ways to satisfy this condition
and correspondingly two types of instantons.
The first one exists when $\xi_2\ne\xi_3$ and only
for $0\le a<1$. It has topology $S^2 \times S^2 - \{pt\}$
and is interpreted as describing the creation of two
oppositely charged dilaton black holes joined by a wormhole.
These ``wormhole'' instantons generalize
the Einstein-Maxwell instanton
discussed in  \garstrom. The reason these instantons
only exist for $0\le a<1$ can be understood by recalling the
thermodynamic behavior
of the dilaton black holes as extremality is
approached: the Hawking temperature, as defined from the
period of $\tau$ in the euclidean section, goes to zero for $0\le a<1$,
approaches a constant for $a=1$ and diverges for $a>1$.
Thus, for small magnetic fields and hence accelerations,
we expect to be able to
match the resultant Unruh temperature and the black hole temperature
by a small perturbation of the black
hole away from extremality only for $0\le a<1$.

The second class of instantons
we mention only for completeness since their interpretation
is obscure. They are defined by
$\xi_2 = \xi_3$ which is equivalent to $r_+A = 2/(3\sqrt{3})$,
and
have topology $S^2 \times R^2$. They are related to the upper limit on
$\hq \hB$ given by \bqlim. Note that for these instantons one does not
have to impose the condition \nonodes\ for regularity.

\subsec{Extremal Instantons}

The wormhole type instantons discussed above were made regular
by the condition that the temperatures of the black hole
and acceleration horizons should be equal.
Gibbons \refs{\gwg} pointed out (for $a=0$) that
there
is another way that the temperatures of the black hole
and acceleration horizons can be equal: that is if the
black hole is extremal. This might seem strange since
extremal dilaton black holes have zero temperature in the
sense that the euclidean time coordinate need not be periodically
identified to obtain a regular geometry. But, of course we
{\it{can}} periodically identify the euclidean time and with any period we
like (just as for flat space). In particular we can choose the
period forced on us by having to eliminate a conical singularity
elsewhere in the spacetime.

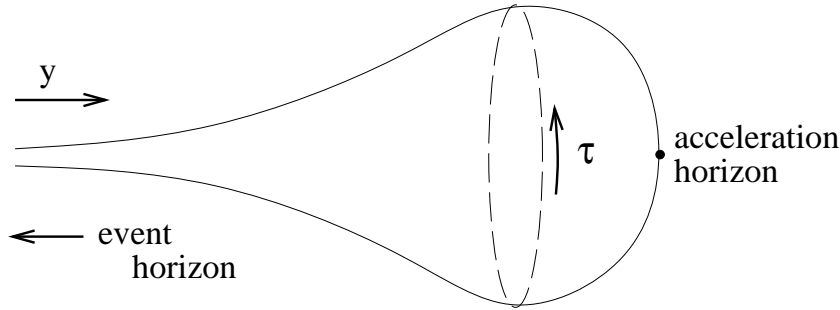
\ifig{\Fig\aeqz}{The $(y,\tau)$ section of the extremal euclidean
$a=0$ solution.  }{pcebh.fig2}{1.8}

{}For $a=0$ the extremal condition $\xi_1=\xi_2$ does indeed lead
to a smooth instanton. The coordinate $y$ lies between
$\xi_2$ and $\xi_3$ in the euclidean section and
we must choose the period of $\tau$ to be again given by \tauper\
in order that there be no conical singularities at
the acceleration horizon, $y=\xi_3$.
We saw in Section 2.2 that the lorentzian
solution near the back hole is just that of an extremal
black hole. The same holds for the euclidean solution.
The horizon $y=\xi_2$ is infinitely far away (in every
direction since every direction is now spacelike) and gives
no restriction on the period of $\tau$. Thus we have
obtained a regular geometry with internal infinities down the
throats of the extremal black holes. The length of the $y=$ constant
circles tends to zero as $y \to \xi_2$, as shown in fig.~2,
but the curvature remains bounded.  Each point in fig.~2 corresponds to a
two sphere, whose area approaches a constant near the event horizon and becomes
large near the acceleration horizon. The figure is slightly misleading
in the vicinity of the
acceleration horizon since  the point corresponding to infinity ($x=y=\xi_3$)
must be removed.  The topology is
$R^2 \times S^2 - \{pt\}$ and the $\tau=0,\Delta\tau/2$
zero momentum slice is a spatial slice of a Melvin universe
with two infinite tubes attached.  The latter is illustrated in
fig.~3.

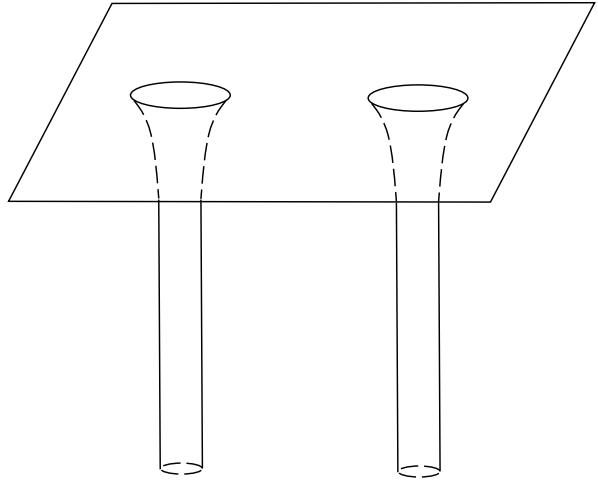
\ifig{\Fig\slice}{The spatial slice $\tau=0,\Delta\tau/2$
through
the instanton solution of fig.~2.  The geometry corresponds to an
asymptotically Melvin region, with two extremal throats attached.  The
solution may be continued to lorentzian signature along this
slice.}{pcebh.fig3}{2.5}

The extremal case $\xi_1=\xi_2$ also gives well defined
instantons for $0<a\le 1$. Although the Einstein
metric has a  singularity, the so called ``total" metric \gm,
$ds_T = e^{2\phi\over a} ds^2$, which is the same as the
string metric for $a=1$, is perfectly regular.
We saw in Section 2.2 that the
metric close to the singularity is that of the extremal
black hole. In the total metric this looks like
\eqn\lindil{
ds_T^2 \propto -d{t'}^2 + \left(1-{\widehat r_+\over r} \right)^{-{4\over
1+a^2}}dr^2
+{\widehat r_+}^2 \left(1-{\widehat r_+\over r} \right)^
{{2(a^2-1)\over 1+a^2}}
d\Omega_2^2
}
{}For $0<a<1$,
the total metric is geodesically complete and the
spatial sections have the form of two asymptotic regions joined by a
wormhole, one region being flat, the other having a deficit solid
angle. Hence the corresponding extremal instantons are regular.
{}For $a=1$ the geometry of the string metric
is that of an infinitely long throat
of constant radius
and thus
the $a=1$ extremal instanton
looks very much like that of the $a=0$
extremal instanton described above (see fig.~4): the topology is the
same, $R^2 \times S^2 - \{pt\}$, and the major difference
is that the proper radius of $y=$constant circles in the $(y,\tau)$
section tends to
a finite limit as $y\rightarrow \xi_2$. The $\tau=0,{\Delta\tau\over2}$
slice resembles the one shown in figure 3.

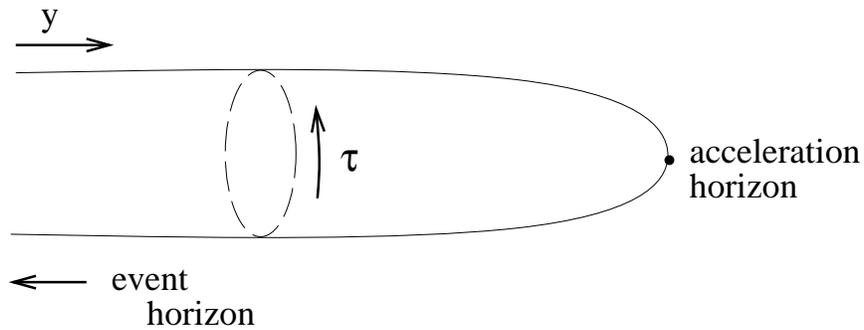
\ifig{\Fig\aeqo}{The $(y,\tau)$ section of the euclidean $a=1$ solution in
the string metric.
}{pcebh.fig4}{1.80}

{}For $a>1$, both the Einstein metric and the
total metric have a  naked singularity in the extremal limit.
It has been argued in \HoWi, however, that these
``black holes" should be interpreted
as elementary particles. The extremal instantons can then be
interpreted as pair creating such objects.
{}For $a=\sqrt{3}$ the instanton can also be understood as
describing the creation of a Kaluza-Klein monopole-anti-monopole
pair (see Section 5).

{}Finally we discuss how these different classes of instantons
fit together in parameter space. The parameters of both
the extremal and the wormhole type
instantons are restricted by the no-nodal
singularities condition \nonodes. The wormhole instantons are further
restricted by \roots\ and this condition is plotted in fig.~5 for
various values of $a$. The extremal instantons satisfy $\xi_1=\xi_2$
and this is also plotted in the figure.  Note that in the limit $a\to1$
the wormhole instantons approach the extremal instantons. It is also
clear from the figure that for $r_\pm A\ll1$ the constraint for both
types of instantons is $r_+=r_-$. The figure also includes the
other class of instantons we discussed when $\xi_2=\xi_3$ or
equivalently $r_+A=2/3{\sqrt 3}$.

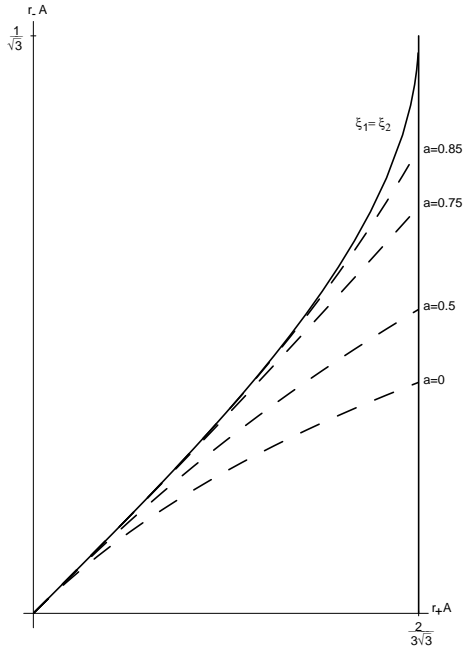
\ifig{\Fig\Params}{Plot in parameter space of the wormhole type
instantons for various $a$ (dashed lines),
the extremal type instantons, $\xi_1=\xi_2$,  and the curve $\xi_2=\xi_3$
($r_+A=2/3{\sqrt 3}$)(solid lines).
}{pcebh.fig5}{3.5}

\subsec{The action}

To leading semiclassical order, the pair production rate of non-extremal or
extremal black holes is given by
$e^{-S_E}$ where $S_E$ is the euclidean action of the corresponding
instanton solutions.
The euclidean action including boundary terms is given by
\eqn\eac{
S_E={1\over 16\pi }\int_V d^4x{\sqrt g}\left[-R+2(\nabla\phi)^2+
e^{-2a\phi}F^2\right]-{1\over 8\pi }\int_{\partial V} d^3x{\sqrt h}K
}
where $h$ is the induced three metric
and $K$ is the trace of the extrinsic curvature of the boundary.
Taking the trace of the metric equation of motion \eom\ yields
$R=2(\nabla\phi)^2$ so the first two terms in the action cancel. The dilaton
equation of motion shows that the third term is a total derivative.
Thus the action of any solution can be recast as a boundary term
\eqn\bterm{
S_E=-{1\over 8\pi }\int_{\partial V}
d^3x{\sqrt h}e^{-{\phi\over a}}\nabla_\mu(
e^{\phi\over a}n^\mu)}
where $n^\mu$ is a unit outward pointing normal to the boundary.
Note that for $a=0$ \bterm\ is still well defined for our solutions
\dernst\ since
 $\lim_{a\to 0}{\phi/a}$ is
finite.

{}For both the wormhole and extremal instantons there is a boundary
at infinity,
$x=y=\xi_3$ which contributes an infinite amount to the action.
However, the action of the background magnetic field solution is itself
infinite.
In the appendix we show how
the infinite  background contribution  is subtracted to
obtain the physical result.
{}For the extremal instantons there is also an additional
boundary down the throats of the black holes i.e. at $y=\xi_2$.
The contribution to the action from this boundary vanishes.

Leaving the details of the calculations to the appendix we
quote the result here.
The action is finite for both types of
instantons and is given by
\eqn\cta{
S_E=
2\pi \hq^2 {\Lambda(\xi_4)(\xi_3-\xi_2)\over \Lambda(\xi_3)(\xi_4-\xi_3)}.
}
Notice that the result is finite for the extremal instantons
despite the infinite throats for $0\le a\le 1$ and despite the
fact that there are
singularities in both the Einstein and the total metric
for $a>1$.
The action can be expressed in terms of the physical charge
$\widehat q$ and magnetic field $\widehat B$ by
expanding out in the parameter $\widehat q\hB$.
The action for the wormhole type instantons is
\eqn\wormacta{
\eqalign{
&S_E=\pi\widehat q^2\[{1\over \widehat q\hB}-{1\over 2}
+\cdots\]
\qquad\qquad a=0\cr
&S_E=\pi\widehat q^2\[{1\over (1+a^2)\widehat q\hB}+{1\over 2}
 +\cdots \]
\qquad 0< a<1\cr}
}
while the action for the extremal type instantons for all $a$ is given by
\eqn\extact{
S_E=\pi\widehat q^2\[{1\over (1+a^2)\widehat q\hB}+{1\over 2}
+\cdots\]
}
where dots denote higher order terms which may be fractional
powers of  $\hq\hB$.
To leading order these all give the Schwinger result, $\pi m^2/
\widehat q\hB$ after using the relation between the
mass and charge of extremal black holes,
$(1+a^2)m^2=\widehat q^2$.

To next-to-leading order, for $a=0$ the action of
the extremal instanton is greater
than the action of the wormhole instanton
by $\pi {\widehat q}^2 = {1\over 4} A$  where $A$ is the area
of the horizon of an extremal black hole of charge $\widehat q$.
In fact, to this order it could also be the area of the horizon
of the wormhole instanton. This difference
is precisely the Bekenstein-Hawking
entropy.  For $0<a<1$ the difference is zero to this order, which
is consistent with the difference being the area of the
horizon of the extremal instanton since that vanishes for $a>0$. The
area of the horizon in the wormhole instanton is non-zero, but  higher order
in $\hq \hB$.

In \GGS\ a comparison was made between the wormhole action
 for $a=0$ and the action of an instanton describing the
creation of a monopole-anti-monopole pair.
It was found that the action of the monopole instanton
was greater than that of the wormhole instanton by
the black hole entropy. Our result thus suggests that,
at least for $a=0$, the extremal black holes behave more like
elementary particles than non-extremal ones.   However, these conclusions
neglect quantum corrections, to which we
now turn.

\newsec{Quantum considerations}

Until now investigation of the solutions has been carried out on the
classical level.  In this section we will discuss quantum corrections, and
see that they have important effects.  Let us begin with some
qualitative observations.
{}First consider the lorentzian solutions with general $m$ and $q$, and
note that an observer travelling on a trajectory a fixed distance
from the black hole will be accelerated\foot{This is of course in addition
to the usual acceleration needed to avoid falling into the black hole
were it static.}
and therefore would observe
acceleration radiation if carrying a detector.  This suggests that we should
describe the black hole as being in contact with this approximately thermal
radiation.  If so, then the black hole would be expected to absorb energy,
and the solution would then not be static.  However, the black hole can
also emit Hawking radiation, and therefore achieve a time-independent
equilibrium state where the emission and absorption rates match.  We have
already seen evidence of this in the wormhole-type
euclidean solutions:  for a regular
solution the periodic identification required for regularity at the
acceleration horizon had to match that required at the black hole horizon.
This corresponds to matching the Unruh and Hawking temperatures, and thus
putting the black hole in equilibrium.  There resulted a condition
determining the mass of the black hole in terms of its charge and the
magnetic field. We will investigate whether similar statements apply
to the extremal case.

 A quantitative  study could be made by canonically
quantizing fluctuations of the fields about the solutions, and computing
the Bogoliubov coefficients. Such a
calculation has recently been done for the case of charged black holes
in de Sitter space \kt.  If the quantum state at ${\cal
I}^-$ is the Melvin analogue of the Minkowski vacuum, then these calculations
should yield the above statement that the black hole is bathed in acceleration
radiation. This is  not in contradiction to our previous observation that the
extremal black
holes have zero proper acceleration.  To see this, consider
quanta of fixed frequency sent toward the black hole from infinity at two
different times.  The surroundings of the black hole at the two different
times at which these quanta reach the black hole are related by a boost.
Therefore in the instantaneous rest frame of the black hole the quanta will
have different frequencies. Thus in general there
is a time-dependent red- or blue-shift between infinity and the black hole,
resulting from the acceleration.
In describing measurements made by observers near the black hole one will
encounter a corresponding Bogoliubov transformation similar to that for flat
space modes in an accelerated frame, along with extra blueshift factors to
account for the gravitational field of the black hole.  This Bogoliubov
transformation will describe the thermal flux of acceleration radiation
into the black hole, and could in principle be used to determine the
quantum stress tensor.

In performing these calculations the forms of the effective potentials for
fluctuations about the black holes are also crucial.
In particular, note that the behavior of fluctuations about
the extremal black holes \refs{\HoWi,\DXBH}
depend critically on the value of $a$.
{}For $a<1$ there are potential barriers outside the black hole, but these
vanish at the horizon and thus permit fluctuations there.  However, for
$a=1$ a barrier develops:  all modes have a non-zero
mass gap.
This is even more pronounced in the case
$a>1$ where the potentials grow as the horizon is
approached.
This suggests that fluctuations are effectively suppressed.

A detailed treatment of these perturbations and of their quantum effects is
rather involved and will be left for future work.  We will instead
make two simplifications of the problem that we believe  preserve the
essential features.  First, rather than
considering the general perturbation of the graviton, Maxwell, and dilaton
fields we will just consider perturbations of a free spectator field $f$,
with action
\eqn\spect{S= -\hf \int d^4 x \sqrt{-g} (\nabla f)^2\ .}
We expect the dynamics of this field to be similar to that of a general
perturbation.\foot{In a theory explicitly including $N$ such fields,
rigorous justification for
neglecting the gravitational and
electromagnetic flucuations can be given in the large $N$ limit.}

It should be noted that it is sometimes appropriate to extend the action
\action\ by explicitly adding other fields that do not have effective
potential barriers.  An example is at $a=1$, where one
finds from string theory massless modes with couplings of the
form \refs{\CGHS,\DXBH}
\eqn\rrcoup{S= -\hf \int d^4 x e^{2\phi}\sqrt{-g} (\nabla f)^2\ .}
The extra coupling to the dilaton effectively removes the mass gap.

The second simplification is to work only in the s-wave sector of the
theory.  This assumption can be justified in a controlled
approximation \refs{\CGHS\DXBH} for the $a=1$ near-extremal solutions with
matter described by \rrcoup.
The reason for this is that in the long
throat of the $a=1$ solution, the potential for the non-spherical modes is
constant and of order $1/q^2$.  This gives an effective mass gap, and if we
consider excitations below this energy we can ignore these higher modes.
In the case of the $a<1$ throats it is less clear that such an
approximation is strictly justified, since then the potential is not
constant and vanishes down the throat.  Nonetheless, we expect that
treatment of the $s$-wave modes should give us a reasonable picture of the
role of quantum effects.

{}For illustration we will focus on the cases $a=0$ and $a=1$.
In both of these there is a
two-dimensional effective action describing the throat region of the black
hole.  The gravitational part of these actions take the form
\eqn\tdacts{\eqalign{S_0=& {1\over4}\int d^2 x \sqrt{-g}\left\{
e^{-2D}\left[R+2(\nabla D)^2\right] + 2-2q^2e^{2D}\right\}\cr
S_1=& { q^2 \over 2} \int d^2 x\sqrt{-g}\left\{e^{-2\phi}\left
[R+4(\nabla\phi)^2
+{1\over 2q^2}\right]\right\} \ }}
where in the former $e^{-D}$ is the radius of the two-sphere cross-section
of the throat, and in the latter $g$ is the two-dimensional reduction of
the total (or string) metric.
These have two-dimensional black hole solutions of the form
\eqn\tdsolns{\eqalign{a=0:\quad& ds^2 = -{(r_+-r_-)^2\over 4q^2}\sinh^2 (x
-x_h)dt^2 + q^2dx^2\ ,\ e^{-D} = q\cr
a=1:\quad& ds^2 = -\tanh^2 (x-x_h)
 dt^2 +8q^2 dx^2\ ,\ e^{2\phi}= {e^{2\hat\phi_0}\over
\cosh^2 (x-x_h)}}}
which correspond to the near-extremal Ernst solutions far down the throat,
up to exponentially small corrections from effectively massive modes.
In the extremal limits $x_h\rightarrow -\infty$ and $\hat\phi_0\rightarrow
\infty$ (for details see \DXBH) and these take the form
\eqn\tdext{\eqalign{a=0:\quad& ds^2 = -{e^{2x}\over q^2} dt^2 +
q^2dx^2\ ,\ e^{-D} = q\cr
a=1:\quad& ds^2 = -dt^2 + 8q^2 dx^2\ ,\ \phi=-x \ .} }
To the actions \tdacts\ must be added the reduced matter actions,
\eqn\mattact{S_f= -\hf \int d^2 x \sqrt{-g} (\nabla f)^2\ ,}
which arise from \spect\ in the case $a=0$ (where the small variations in
$D$ are neglected) and \rrcoup\ in the case $a=1$.
(Here $f$ has been rescaled by a $q$ dependent constant.)

By working with the two-dimensional theory we can
compute the  expectation value of the
quantum stress tensor using the connection with the
conformal anomaly \refs{\ChFu,\CGHS}.   Transforming to conformal coordinates,
\eqn\confcoord{ds^2 =  e^{2\rho}(-dt^2 + dy^2)=
 - e^{2\rho} d\sigma^+ d\sigma^-\ ,}
this takes the form
\eqn\stresst{\eqalign{T^f_{+-}& =-{1 \over 12} \partial_+\partial_-\rho\
,\cr T^f_{++}& = -{1 \over 12} \left(
 \partial_+\rho \partial_+\rho - \partial^2_+\rho +
t_+(\sigma^+)\right)\ ,\cr
          T^f_{--}& =-{1 \over 12} \left(
  \partial_-\rho\partial_-\rho - \partial^2_-\rho +
t_-(\sigma^-)\right)\ ,}}
where $t_+$ and $t_-$ are to be determined by the boundary conditions.
The leading quantum corrections to the solutions can be found by including
these on the right hand side of Einstein's equations.
If we are looking for static solutions, we should demand that $t_+=t_-=t_0$
is a constant.

The boundary conditions of the two-dimensional theory are to be determined
by matching correctly onto the four-dimensional theory in the region where
the throat matches onto the asymptotic region.  One obvious possibility is
that the boundary conditions be chosen so that the two-dimensional quantum
state is the vacuum annihilated by the positive frequency modes in the time
variable $t$.
This implies $t_0=0$.  However, this is not a physically realizable state.
To see this, note that in the context of the full four-dimensional theory,
the state is that annihilated by the positive frequency modes defined with
respect to the Killing vector.  Asymptotically this Killing vector
corresponds to the boost symmetry of \blip, and
thus the state
 tends to a
Rindler-like vacuum at infinity.
As seen by an observer at rest with respect to the
magnetic field, this vacuum has infinite stress tensor, and thus
becomes singular, on the acceleration horizon.

A more appropriate state at ${\cal I}^-$ is the vacuum as defined by an
observer stationary with respect to the asymptotic Melvin solution.
{}From our earlier arguments, this state will not appear to be vacuum for an
observer near the horizon.  There will be particles arising from the
non-trivial Bogoliubov transformation, and
a flux of acceleration
radiation is expected in the vicinity of the black hole.  This corresponds
to $t_0\neq0$; in
general one would expect $t_0$ to be proportional to the acceleration of
the black hole.  To
determine the actual value of $t_0$ requires knowing the details of the
matching, and this is difficult.  We can however see that $t_0$ will have a
major effect on the solution.  Consider for example $a=1$.  In this case
the string metric of the classical solution is perfectly regular, and
 tends to a product of the linear dilaton vacuum and the round
two-sphere down the throat.
However,
equations
from \tdacts,\mattact, with the quantum corrections \stresst, have been
investigated both numerically and analytically in
\refs{\BGHS\SuTh-\Hawktd}.
There
it was found that for general $t_0$ the static
solutions have singular horizons.  These result from a non-vanishing stress
tensor penetrating into the region where the theory is effectively strongly
coupled.  The exception to this is when
the ingoing flux $t_+$ matches the outgoing flux due to Hawking radiation.
This could happen only at a large definite acceleration with $r_+
A$ of order 1.

The story for $a=0$ is similar.  The static equations were investigated in
\refs{\Triv}.  There it was found that the equations are singular for all
$t_0$.  However, for $t_0=0$ the singularity is mild and quantum corrected
solutions were found.  In contrast, at $t_0\neq0$ more serious
singularities arise.  This can be readily confirmed by writing
the static equations in coordinates regular at the horizon, similar to the
discussion
in \refs{\BGHS}.

The preceding arguments are also expected to generalize to $0<a<1$.  However,
note that they do not apply to $a>1$, as in this case the growing
potentials mean that the action \mattact\ is not a good approximation near
the horizon -- fluctuations are effectively suppressed in this region.

We therefore conclude that for $0\leq a< 1$, or for $a=1$ with matter given
by \rrcoup, quantum corrections become large and
the semiclassical approximation fails near the black hole.
The detailed construction of the fully quantum-mechanical solutions
is therefore unknown and may depend on new physics.
There should certainly exist {\it some} sensible solutions that
closely resemble the classical solutions away from this region -- one
certainly
hopes to be able to give a physical description of the equilibrium state of
a charged black hole
in a background electromagnetic field.  It could be that
the physical equilibrium solutions
correspond to the lorentzian version of the sub-extremal solutions
of \refs{\garstrom,\GGS,\DGKT}, or it could be that
there are different physical
solutions corresponding to the quantum corrected extremal black holes of
this paper.  Note that
although our arguments have been made with the $f$ fields, we expect this
instability to quantum corrections to persist with more general
perturbations for $a<1$.  However, for $a=1$ without the fields in \rrcoup\
this argument no longer applies.

Similar considerations apply to the euclidean solutions.  Indeed, the
euclidean solutions should be time-symmetric on the slice of constant euclidean
time along which we cut them to match to the lorentzian solutions.  As
before, the role of quantum corrections can be inferred from the one-loop
action of the matter field $f$.  As in the lorentzian case, the stress
tensor for minimally-coupled s-wave matter
can be explicitly computed in conformal gauge, and the result is the
analytic continuation of  \stresst.  Here one again expects $t_0$ to be
non-zero when the four-dimensional and two-dimensional solutions are
matched.  This has the unfortunate consequence of yielding large
corrections to the equations of motion in the vicinity of the horizon --
the back-reaction becomes strong and the
semiclassical approximation breaks down.  This means that without
leaving the semiclassical approximation the structure of the pair-produced
objects cannot be determined near the horizon.  It is plausible
that once quantum corrections are included for $a<1$
the corresponding
geometry is similar to the black holes connected by
Wheeler wormholes of \refs{\garstrom,\GGS,\DGKT} except in
the immediate vicinity of the horizon.
{}Furthermore, the rate
cannot be calculated and may depend on new physics, and in particular on
the existence of fundamental charge in the theory.\foot{This is
in contrast to the case of Wheeler wormholes where quantum corrections are
not necessarily large \Cilar\ and fundamental charge is not required.}
It is, however, reasonable to expect
this production rate
 to be non-zero.    One reason for believing this is
that the objects we are considering are clearly in a different
topological class from wormholes -- the geometries are not
connected through the throat -- and it seems unlikely that the production
rate would be zero in this sector.  This belief is reinforced in the case
$a=\sqrt{3}$ where, as we will describe, there is no infinite
throat, the fluctuations do not make large contributions, and the euclidean
solution describes pair production of Kaluza-Klein monopoles.  It is
plausible that this type of production extends to $a\leq 1$.

It is also worth commenting on the  issue of production rates for
\RN\ black holes. If information is not lost in black hole
formation and evaporation and
 it does not escape in Hawking radiation, this implies that a \RN\
black hole has an infinite number of states,\foot{This has been
particularly convincingly argued in \refs{\StTr}, using
semiclassical techniques.} and na\"\i ve effective field
theory reasoning would then imply an infinite production rate.  A possible
resolution to this was suggested in \refs{\Cilar}, building on
ideas in \refs{\BaOl,\BOS}:  although \RN\ black
holes do have infinite states, not all such states are produced by the
euclidean instantons.  Ref.~\Cilar\ argued this for the case where the
black holes are connected by a wormhole, although
similar reasoning applies here as well.
The basic point is that fluctuations of the infinite number of states near
the black hole lead to a large quantum stress tensor and therefore a large
back-reaction.  Indeed, when
computing the
amplitude for any process involving black holes, contributions of these states
are summarized in the functional integral over fields in the black hole
background; for example, in the case of $f$-states,
\eqn\infstates{\int \cald f e^{iS[f]}\ .}
When continued to euclidean signature, this expression might at first sight
be expected to include an overall infinite factor counting these
states.
However, as discussed above, the quantum stress tensor
derived from this functional
integral becomes large near the horizon, precisely because of these
infinite states, and this signals a breakdown of the
semiclassical approximation.  Although this means that the rate cannot be
calculated to find whether it is finite or infinite, it is
also an indicator that the na\"\i ve effective
field theory logic is breaking down.  The non-trivial dynamical role of
this functional integral is in contrast to a rate of the form
\eqn\naivr{\Gamma\sim N e^{-S_{\rm instanton}}}
that one would expect from an instanton that produced $N\rightarrow\infty$
states with
comparable amplitudes.  The failure to obtain a na\"\i vely infinite
rate of the form \naivr\
can be viewed as a strong suggestion that a correct quantum calculation
would in fact yield a finite answer, resolving the problem of infinite pair
production.  Whether such a result can be obtained in a type of effective
theory \refs{\CBHR}
or lies entirely outside the domain of effective theory remains to
be seen.

\newsec{The Kaluza-Klein Case}

As we have remarked several times,
the value $a = \sqrt 3$ is of special interest since in this case the
action $S$ is equivalent to Kaluza-Klein theory. In other words, if
$g_{\mu\nu}, A_\mu, \phi$ are  an extremum of $S$ with $a = \sqrt 3$,
then one can construct a five dimensional solution of the vacuum Einstein
equations by
\eqn\fivmet{ ds^2 = e^{-4\phi/\sqrt 3}( dx_5 + 2A_\mu dx^\mu)^2
     +e^{2\phi/\sqrt 3} g_{\mu\nu} dx^\mu dx^\nu \ . }
Since the fields do not depend on the fifth coordinate $x_5$, this
solution always has at least one
translational symmetry. In this section we will explore
the five dimensional vacuum spacetimes associated with the
dilaton Ernst
solutions \dernst\ in both the lorentzian and euclidean contexts.

We begin with the static magnetically charged black hole \dbhs. Setting
$ a = \sqrt 3$  and substituting the fields into \fivmet\
yields the following five dimensional metric \kkbhs
\eqn\kkbh{\eqalign{
 ds^2 =& -\(1-{r_+ \over r}\) \(1-{r_- \over r}\)^{-1} dt^2 +
  \(1-{r_+ \over r}\)^{-1} dr^2 \cr +& \(1-{r_- \over r}\)
  \[dx_5 + 2q (1-\cos \theta )d\vp \]^2 + r^2 \(1-{r_- \over r}\) d\Omega^2
  }}
This spacetime has a horizon at $r=r_+$ and a singularity at $r=r_-$.
In the extremal limit, $r_+ = r_-$, the metric becomes
\eqn\kkmon{ds^2 = -dt^2 + {dr^2\over\(1-{r_+ \over r}\)}  + \(1-{r_+ \over r}\)
  \[dx_5 + 2q (1-\cos \theta ) d\vp \]^2 + r^2 \(1-{r_+ \over r}\)
d\Omega^2\ .}
The horizon is no longer present. There appears to be a singularity at
$r=r_+$, but if we set $\rho = 2 r_+^{1/2}
(r-r_+)^{1/2},$ then near  $r=r_+$ the metric takes the form
\eqn\kkreg{ ds^2 = -dt^2 + d\rho^2 + {\rho^2\over 4} \[ ( d\psi +(1-\cos\theta)
 d\vp )^2 + d\theta^2 + \sin^2 \theta d\vp^2 \]   }
 where we have set $\psi = x_5/2q$ and used the fact \mass\
that $4q^2 = r_+^2$.
If $\psi$ is periodic with period $4\pi$, then
the quantity in brackets is just the metric on a three sphere
of radius two, expressed in terms of Euler angles.
So the solution \kkmon\ is globally regular and free of singularities provided
$x_5$ has period $8\pi q$. It is
the Sorkin-Gross-Perry Kaluza-Klein monopole \refs{\rafKK,\grossperry}.
At large
$r$ it asymptotically
approaches the product of $S^1$ and four dimensional Minkowski space.
Globally, it is
the product of time and the Taub-NUT instanton.
Its topology
is simply $R^5$.

Next we turn to the background magnetic field solution \dmelv. Setting
$ a = \sqrt 3$ and substituting into \fivmet\ yields
\eqn\kkmel{ds^2 = -dt^2 +dz^2 + d\rho^2 +
    \Lambda \(dx_5 + {B \rho^2 d\vp \over\Lambda}\)^2  +
    {\rho^2 d\vp^2\over \Lambda} }
where $\Lambda = 1+ B^2 \rho^2$. This metric is actually flat. It can be
simplified to
\eqn\kkflat{ ds^2 = -dt^2 +dz^2 +d\rho^2 +dx_5^2 +
	        \rho^2 (d\vp + B dx_5)^2  }
How can a flat five dimensional space produce nontrivial
four dimensional fields? The point is that one is reducing to four dimensions
not along the trivial translation in the fifth direction, but rather along
a linear combination of that translation and a rotation \DGKT. This is why
$\vp$ is shifted in \kkflat.
The result is not, in general, globally equivalent to the standard
Kaluza-Klein vacuum. For almost all values of $B$, even though the
metric on the 2D torus of constant $t$, $z$ and $\rho\neq 0$ in
\kkflat\ is flat, it is globally inequivalent to the metric with $B=0$.
Only if the period of $Bx_5$ is an integer multiple of $2\pi$ are the
metrics equivalent.
In this case one can start with (globally) the same  five dimensional
spacetime, and reduce to
obtain either the  magnetic field or the trivial four dimensional solution.
However, in general, the five dimensional flat
space \kkflat\ is identified in a way which
is different from the Kaluza-Klein vacuum.

{}Finally we turn to the dilaton Ernst solution. The five dimensional
metric is free of the fractional powers present in the four dimensional
solution. It is most conveniently
described in terms of functions $\bar F, \bar G$ which are simplified
versions of the functions $F,G$ which appeared in the four dimensional
metric:
\eqn\kkfns{\eqalign{
&\bar F(\xi)=(1+r_-A\xi)\cr
&\bar G(\xi)=\left[1-\xi^2-r_+A\xi^3\right] \ .
}}
Substituting the solution \dernst\ with $a=\sqrt 3$ into \fivmet\ yields
\eqn\final{\eqalign{
ds^2 =& {e^{-4\phi_0\over {\sqrt 3}}
\Lambda \bar F(y) \over \bar F(x) }\( dx_5 + 2A_\vp d\vp\)^2\cr +
  {e^{2\phi_0\over {\sqrt 3}}
\over A^2 (x-y)^2} &\[ \bar F(x)^2\({\G (y) dt^2 \over \F (y)}
  -{dy^2 \over \G(y)} \)  + \F (y)\( {\F (x) dx^2 \over \G (x)}
  +{\G (x) d\vp^2 \over \Lambda} \) \right] \cr }}
where $\Lambda$ and $A_\vp$ are given by
\eqn\kkla{\eqalign{
A_\varphi=&-{e^{{\sqrt 3}\phi_0}\over 2B\Lambda}(1+2Bqx)
+k\cr
\Lambda=&(1+ 2Bqx)^2+{B^2 \G(x)\F(x)\over A^2(x-y)^2} \ .
 }}

Since $\G$ is just the cubic part of $G$, its roots are the same as
in our earlier discussion,  $\xi_2, \xi_3, \xi_4$, and
$\xi_1 = -{1\over r_-A}$ is the root of $\F$. The non-extremal case,
$\xi_1 < \xi_2$,  has a structure similar to the four dimensional solution.
There is an acceleration horizon at $y=\xi_3$, a
black hole horizon at $y=\xi_2$ and a singularity
at $y=\xi_1$. The ranges of the coordinates are $\xi_1 < y < x$ and
$\xi_3 \le x \le \xi_4$. In the extremal limit $\xi_1 = \xi_2$,
the situation is different. One can see immediately from \final\
that in this case $g_{tt}$ approaches a constant as $y \rightarrow \xi_2$.
In fact, we showed in Section 2.2 that the extremal dilaton Ernst solution
approaches the extremal static black hole as $y \rightarrow \xi_2$ with a
constant shift in the dilaton. If the dilaton was not shifted, we could
use the relation between the extreme black hole and the monopole to immediately
conclude that
the metric \final\ is nonsingular
at $y=\xi_2$  provided we identify $x_5$ with period $8\pi \widehat q$
where $\widehat q$ is the physical charge. It turns out
that the  constant shift
in the dilaton does not affect this conclusion. One way to see this is
to rewrite the five dimensional metric in the form
\eqn\fivmeta{ ds^2 = e^{2\phi/\sqrt 3}\[
     (e^{-\sqrt 3 \phi} dx_5 + 2 e^{-\sqrt 3 \phi} A_\mu dx^\mu)^2
     + g_{\mu\nu} dx^\mu dx^\nu \] \ . }
To satisfy the field equations \eom, when a constant is added to $\phi$,
the gauge field
must also be rescaled  in such a way that $e^{-\sqrt 3 \phi} A_\mu $
is invariant. So if we start with the metric \kkmon\ with periodicity
$8\pi q$ for $x_5$, and add a constant $\phi_0$ to $\phi$, then regularity
requires  $e^{-\sqrt 3 \phi_0} x_5$ to have the same period $8\pi q$.
Thus $x_5$ has period $8\pi q e^{\sqrt 3 \phi_0}$. But $q e^{\sqrt 3 \phi_0}$
is just the physical charge after the dilaton has been shifted.

The solution \final\ can thus be viewed as describing a pair of oppositely
charged Kaluza-Klein monopoles accelerating in a background magnetic field.
This is not strictly accurate since the origin of each monopole
is not accelerating:  one can show
that the worldline $y=\xi_2$ is a geodesic. (This is analogous to the fact
that the horizon of the extremal Ernst solution is not accelerating, which
was discussed in Section 2.2.)
However, all points away
from the center are accelerating, and the monopole is not spherically
symmetric.

We showed in Section 2.1 that the dilaton Ernst solution approaches the
background magnetic field at large distances. Since the five dimensional
metric associated with this background field is flat, we conclude
that \final\ is asymptotically flat. We have seen that even though the
magnetic field solution is flat,
it is generally not equivalent to the standard Kaluza-Klein vacuum.
It will be globally equivalent
only if the period
of $x_5 $ is an integer multiple of $2\pi/\widehat B$. But in \final,
the period of $x_5 $ is fixed by regularity at the center of the monopole
to be $8\pi \widehat q$. Thus \final\ approaches the standard Kaluza-Klein
vacuum only if $\widehat q\widehat B =n/4$ for some integer $n$. However as
discussed in Section 2.2, there
is an upper limit on $\widehat q\widehat B$ coming from the fact that
$\xi_2 <\xi_3$.
 Setting $a^2 = 3$ in eq.~\bqlim\
yields $\widehat q\widehat B <1/4$. So the asymptotic region is never
equivalent to the standard Kaluza-Klein vacuum, but instead includes an
identification  involving a rotation as well as a translation.

Even with the nontrivial identifications at infinity, it is interesting
that \final\ is a globally regular, nontrivial, asymptotically flat
solution of the five dimensional vacuum Einstein equations. It is also
dynamical
in the sense that the Killing vector $\p /\p t$ is not asymptotically a
time translation, and is spacelike in some regions. This is
difficult to achieve in four dimensions. In fact, to the best of
our knowledge, there is no analogous solution known in that case. However
it is easier to achieve in five dimensions. Another solution of this type
was previously found by Witten \refs{\Witt}. By taking the five dimensional
Schwarzschild solution and analytically continuing in both $t$ and $\theta$
he obtained
\eqn\kkschw{ ds^2 = -r^2 dt^2 + \(1- {2M\over r^2}\)^{-1}dr^2
  + \(1- {2M\over r^2}\) d\chi^2 + r^2 \cosh ^2 t d\Omega^2}
This solution describes a bubble undergoing uniformly accelerated expansion in
spacetime. Like \final, it is nonsingular, dynamical, and asymptotically flat.
In fact it has another feature in common with \final. Before describing
it let us recall that the positive energy theorem does not hold in
Kaluza-Klein theory if surfaces of different topology are allowed: there
are regular initial data with negative energy \refs{\brill}.  Witten's bubble
\kkschw\ has zero ADM energy and can be interpreted as a possible
outcome for the decay of the Kaluza-Klein vacuum. Our solution \final\
also has zero ADM energy. This follows from the fact that there is a
boost symmetry, and a timelike ADM energy-momentum vector would not be
invariant under such a symmetry, and corresponds to the statement that the
solution has the same energy as the corresponding Kaluza-Klein Melvin
solution \kkflat.
One can thus view \final\ as a possible outcome for the decay of this
solution.

The corresponding instanton,
obtained by  replacing $t$ with $i\tau$
in \final, can be viewed as creating a pair
of Kaluza-Klein monopoles.
The metric is positive definite if the coordinate $y$ is
restricted to lie in the range $\xi_2 \le y \le \xi_3$.
The period of $\tau$ is fixed by regularity at the acceleration horizon
$y=\xi_3$. There is no restriction on the period coming from regularity
at $y=\xi_2$ since the metric approaches \kkmon\ in this region.

The topology of this instanton is $S^5-S^1$. To see this consider
slicing the manifold into two pieces along $y={\xi_3-\xi_2\over 2}$,
say. The piece that contains $y=\xi_2$ has topology $D^4\times S^1$,
with the $S^1$ being the euclidean time. The piece containing
$y=\xi_3$ has topology $S^3\times D^2 -S^1$: the $D^2$ comes from
the $y,\tau$ part of the metric and the subtracted $S^1$ is $x=y=\xi_3$.
The instanton is obtained by gluing these pieces along their common
boundary $S^3\times S^1$ by the obvious diffeomorphism. Using the fact that
$S^5=\partial(D^6)=\partial(D^4\times D^2)=D^4\times S^1\cup_{S^3\times S^1}
S^3\times D^2$ we deduce that the topology of the instanton is indeed
$S^5-S^1$.
We can also show that the topology of the zero momentum slice
$\tau=0,\Delta\tau/2$ is given by $S^4-S^1$. Consider slicing this four
manifold along $y={\xi_3-\xi_2\over 2}$ as before.
The piece that contains $y=\xi_2$ is simply two copies of $D^4$,
while the piece containing
$y=\xi_3$ has topology $S^3\times D^1 -S^1$. Gluing these along the
common boundary $S^3\cup S^3$ gives $S^4-S^1$.

The exact
action
for this instanton is given by \cta\ with $a^2 = 3$.  Expanded in powers of
$\widehat q\widehat B$ the result is
\eqn\extacct{
S_E=\pi\widehat q^2\[{1\over 4\widehat q\hB}+{1\over 2}
 +\cdots\] }
The semi-classical pair creation rate is thus $\Gamma = e^{-S_E}$.
As discussed in the previous section, unlike the
situation for $a=0$ or $a=1$ extremal instantons, the quantum
corrections should remain small and the instanton approximation should
be valid. This is because fluctuations near $y=\xi_2$ should be suppressed
by the large potential barriers, or equivalently from the regularity of the
five-dimensional solution.  Indeed, for weak magnetic
fields and large
charge, the curvature is small everywhere and the
quantum corrections will be small.

\newsec{Discussion}

As we have seen, the extremal limit of the dilaton Ernst solutions found in
\DGKT\ have a number of
interesting properties. These include the fact that the lorentzian solutions,
near the horizon, reduce exactly to the static dilaton black holes.
Analytic continuation yields a finite action instanton which describes the
pair creation of  Kaluza-Klein monopoles when $a=\sqrt 3$, or
extremal black holes for $0 \le a \le 1$. These extremal black holes
contain infinite throats (in an appropriate metric)
and are topologically different from the
wormhole originally discussed in \garstrom\ and generalized in \DGKT.
We have also considered possible quantum corrections to this
leading order semi-classical approximation, and found that in certain
cases they can become large. These corrections can  affect both the
geometry down the throat, and the physical pair creation rate.

Many open problems remain. Some have been mentioned earlier, and include
a better understanding of the limit on $\widehat q\widehat B$, \bqlim,
and the fact that
the difference of the actions for the wormhole and extremal instantons for
$a=0$
is the Bekenstein-Hawking entropy. One of the most important is to
develop a better understanding of the quantum corrections to the instanton
approximation and their
effects on the geometry and pair creation rate.
It is particularly important to understand the
calulation of the rate,
as a finite answer may indicate that such
black holes serve as a model for black hole
remnants \refs{\CGHS,\BDDO,\DXBH,\BaOl,\BOS,\StTr,\Cilar}.
It is notable
that because of the higher order quantum effects,
one does not immediately recover the na\"\i ve estimate of an infinite
rate arising from the infinite number of states. A better understanding
of these corrections will also help to resolve the question of
whether an infinite volume of space can really be created in a finite
amount of time. If so, there would appear to be problems with causality,
unless the state down the throats were fixed uniquely.

Another interesting problem is to understand the behavior of charged black
holes when the background fields are turned on or off in a finite time.
Suppose one starts with an extremal black hole and slowly turns on a magnetic
field. Will it stay extremal? We have seen that the solution right near the
horizon is independent of the magnetic field. But there will  certainly
be an effect
on the solution farther from the black hole which can propagate toward the
horizon. The outcome seems to depend on $a$. The large potential barriers
\HoWi\
for $a>1$, or for $a=1$ with any matter other than \rrcoup,  indicate
that the perturbations never reach the horizon. The black holes stay extremal.
In particular, a Kaluza-Klein monopole should not turn into a magnetically
charged black hole if a magnetic field is turned on. On the other hand,
for $a<1$, the potential barriers
vanish at the horizon.  This together with the second law of
black hole thermodynamics strongly suggests that any time-dependent
perturbation will raise the mass of the black hole away from
extremality.  (One could
perhaps produce an extremal accelerating
black hole  with $a<1$ by first accelerating any charged black hole
and then adding charged particles with $q>m$.)

This dependence on $a$ is further supported by considerations of black hole
thermodynamics. Recall that for the static dilaton black holes, the Hawking
temperature vanishes in the extremal limit only for $a < 1$. It reaches
a constant for $a=1$ and diverges for $a>1$. Thus, if one turns on a
weak magnetic field, one could match the Unruh temperature of the
acceleration with the Hawking temperature of a slightly non-extremal
black hole only if $a<1$. We have already encountered the euclidean
analogue of this statement in Section 3. Wormhole type instantons require
the same matching of temperatures and hence only exist for $a<1$.
A particular puzzling case is
$a=1$ with the special matter \rrcoup. Since the potential barriers vanish
near the horizon, one might expect the accelerated black hole to become
slightly nonextremal, but the temperatures cannot be matched in this case.
It is not clear what the equilibrium solution is.

In the quantum theory, other important questions are related to observations
by observers at infinity and the issue of energy balance.
Suppose, as suggested above,  that an accelerating
black hole is continually emitting Hawking radiation to stay in equilibrium
with the
acceleration radiation that it absorbs.  One would expect this Hawking
radiation to
be observed at infinity.  But the stationary observer does not see the
acceleration radiation. Where does she say that the energy is coming
from to maintain equilibrium?
It may be that,
as in the case of an accelerating
charge, to describe energy balance requires understanding the details
of the switching on and off of the background field \refs{\QEP}.  One more
intriguing question is whether the black hole continually and indefinitely
swallows quantum information in this process,
producing an arbitrary amount of entropy
in the outgoing state.

We close with one final issue. There is presumably a rotating analogue of
the dilaton Ernst solution \dernst. (A rotating analogue of the
$a=0$ C-metric is already known \refs{\roternst}.) It is likely that
this could be analytically
continued to discuss pair creation of rotating black holes. There is reason
to believe that this will provide a wormhole instanton for $a=1$. One
piece of evidence comes from \BOS, where  an approximate
wormhole instanton was constructed
which includes rotation. Another comes from the
fact that for the rotating black hole with $a=1$, the
Hawking temperature goes to zero in the extremal limit whenever the angular
momentum is non-zero \rotbh. Thus one could match the Unruh temperature at the
acceleration horizon by a slightly non-extremal rotating black hole.

\bigskip\centerline{\bf Acknowledgements}\nobreak
We would like to  thank
T. Banks, M. Britton, J. Friedman, D. Garfinkle, R. Geroch, J. Harvey,
D. Kastor, M.
O'Loughlin, E. Martinec, R. Sorkin, A. Strominger, R. Wald and
S. Weinberger for useful conversations.
{}FD and GTH are supported in part by NSF grant PHY-9008502.
JPG is supported by a grant from the
Mathematical Discipline Center of the Department of Mathematics,
University of Chicago.  SBG is supported in part by NSF PYI grant
PHY-9157463 and by DOE grant DOE-91ER40618.

\appendix{A}{Calculating the action}
We present here more details of the calculation of the action of
the instantons discussed in Section 3.3. As shown there, the
action of any solution can be reduced to a surface term
\eqn\bbterm{
S_E=-{1\over 8\pi }\int_{\partial V}
d^3x{\sqrt h}e^{-{\phi\over a}}\nabla_\mu(
e^{\phi\over a}n^\mu) \ .}
{}For both the extremal and wormhole instantons there is a boundary
at infinity, $x=y=\xi_3$. We will evaluate the action on the surface
$y=x-\epsilon$ and then take the limit $\epsilon\to 0$. This will
enable us to properly subtract off the infinite contribution of the
background magnetic field
spacetime.

Performing the trivial integrals over $\vp$ and $\tau$ yields
\eqn\acty{
S_E=-{1\over 8\pi }\Delta\varphi\Delta\tau\int_{\xi_3}^{\xi_3+\epsilon}dx
{\sqrt h}e^{-\phi\over a}{1\over \sqrt g}\p_\mu(e^{\phi\over a}
{\sqrt g}n^\mu)\big|_{y=x-\epsilon}
}
where $\Delta\vp$ is the range of
$\vp$,
\phiperiod, and $\Delta\tau$ is the range of $\tau$ which is given by \tauper\
for both types of instantons.
The individual terms that appear in this integral are given as follows.
The unit outward pointing normal to the surface $y=x-\epsilon$
has components
\eqn\norm{
\eqalign{
&n^y=-{A (x-y) G(y) F(y)^{1\over 2}\over
\Lambda^{1\over 1+a^2}F(x)^{1\over 2}[F(x)G(x)-F(y)G(y)]^{1\over 2}}\cr
&n^x=-{A (x-y) G(x) F(x)^{{1\over 2}}\over
\Lambda^{1\over 1+a^2}F(y)^{{1\over 2}}[F(x)G(x)-F(y)G(y)]^{1\over 2}}\ .
}
}
The induced three metric $h$ on the surface can be constructed and used
to obtain
\eqn\three{
{\sqrt h}=
A^{-3}\epsilon^{-3} \Lambda^{1\over 1+a^2}
{}F(x)^{{1\over 2}}F(y)^{1\over 2}
[F(x)G(x)-F(y)G(y)]^{1\over 2} \ .
}

Expanding in powers of $\epsilon$ and integrating, \acty\ becomes
\eqn\actbec{S_E={\pi L^{2}\over  A^2G^\prime(\xi_3)}
\left(-{3F(\xi_3)\over  \epsilon}
+{F^\prime(\xi_3)\over  a^2 }
+\calO(\epsilon)\right) \ .}
The first term diverges as $\epsilon \to 0$.
In terms of the asymptotic Melvin coordinates $\rho, \zeta$ introduced in
\coordch, this term is simply  $-3\pi
( \rho^2 + \zeta^2)/4$. It is independent of the black hole charge
and is precisely the action of the euclidean analogue of the dilaton
Melvin solution \blip. Notice that the dependence on $\widehat B$ cancels.
Since we are only interested in the difference
between the euclidean Ernst and Melvin actions
we subtract the leading term in \actbec.

{}For the extremal instantons there is an
additional boundary down the throats of the black holes $y=\xi_2$.
A similar calculation to the above, but much simpler, shows that this
boundary does not contribute to the action and hence the action of the
extremal instantons is also given by \actbec.
Thus the action for both types of instantons is finite and given by
\eqn\level{
\eqalign{S_E=&
{\pi L^2F^\prime(\xi_3)
\over a^2 A^2 G^\prime(\xi_3)}\cr
=&2\pi {\widehat q}^2 {\Lambda(\xi_4)(\xi_3-\xi_2)\over
\Lambda(\xi_3)(\xi_4-\xi_3)}\cr}
}
where we have used the expression \pchge\ for the physical
charge $\widehat q$.

For general $a$ we have not been able to express $S_E$ exactly in terms of the
physical charge, $\hq$, and physical value of the background magnetic
field, $\hB$ given by \blob.
Using \hatbq, one can  however establish that in the case of the $a=0$
wormhole instantons, \level\ agrees with the exact result obtained in
\GGS, namely
\eqn\compare{
S_E=4\pi {\hq}^2 {(1-\hq\hB)^2\over 1-(1-\hq\hB)^4}\ .
}

A necessary condition for the instanton approximation to be valid is that
$\hq\hB$ be small. Thus we would like to expand $S_E$ in
$\hq\hB$.
We will expand the quantities that appear in \level\ in terms
of $\delta={1\over\xi_2}$. Then we expand $\hq\hB$ in terms of
$\delta$ and invert the relation. Finally we will use these
expansions to find the leading order behavior of $S_E$ and the next
order correction. The calculation will be different in the two cases
of the wormhole and extremal instantons since the conditions on the
parameters differ.

{}First we give some relations which hold for both the wormhole and
extremal cases.  From the definition of the
roots of the cubic in the function $G$ we deduce
\eqn\cond{
\eqalign{&\xi_2\xi_3\xi_4={1\over r_+A}\cr
&\xi_2\xi_3+\xi_3\xi_4+\xi_2\xi_4=0\cr
&\xi_2+\xi_3+\xi_4=-{1\over r_+A}\cr
}}
and hence that
\eqn\xis{
\eqalign{&\xi_3=-1+{1\over 2}\delta-{5\over  8}\delta^2+\cdots\cr
&\xi_4=1+{1\over 2}\delta+{5\over  8}\delta^2+\cdots\ .
}
}
\subsec{Wormhole}
We first consider the wormhole instantons for $0\le a< 1$ defined
by the constraints \nonodes, \regular. From \roots\ and \xis,
we deduce
\eqn\xione{
\xi_1={1\over \delta}\[1+(-2\delta)^{1+a^2\over 1-a^2}+\cdots\]\ .
}
Next using \nonodes\ and \hatbq
\eqn\blah{
\eqalign{&\widehat q\hB=-\delta\(1+{3\over 2}\delta+\cdots\)\qquad a=0\cr
(1+a^2)&\widehat q\hB=-\delta\(1+{1\over 2}\delta+\cdots\)\qquad0<a<1\ .}
}
Expressing $\delta$ in terms of $\widehat q\hB$ and using \level\ we
obtain the action for the wormhole type instantons
\eqn\wormact{
\eqalign{
&S_E=\pi\widehat q^2\[{1\over \widehat q\hB}-{1\over 2}
+\cdots\]
\qquad a=0\cr
&S_E=\pi\widehat q^2\[{1\over(1+a^2) \widehat q\hB}+{1\over 2}
+\cdots\]
\qquad 0< a<1\  .
}}
\subsec{Extremal}
We now turn to the extremal instantons defined by $\xi_1=\xi_2$
and \cons. Using \xis\ and \cons\  we first deduce that
\eqn\extcond{
(1+a^2) qB=-\delta+\calO(\delta^3)
}
and from \hatbq\ that
\eqn\bblob{
(1+a^2) \widehat q\hB=-\delta\(1+{1\over
2}\delta\)+\calO(\delta^3)\ .
}
Expressing $\delta$ in terms of $\widehat q\hB$ and using \level\ we
obtain the action for the extremal type instantons for all $a$
\eqn\extactt{
S_E=\pi\widehat q^2\[{1\over(1+a^2) \widehat q\hB}+{1\over 2}
 +\cdots\]\ .
}

The mass of a static extremal black hole is given by \mass\ with
$r_+=r_-$, so $ m = {\hq\over \sqrt{1+a^2}}$. Thus we see that the
leading order term in the action in all cases is
$S_{0} = {\pi m^2\over \hq\hB}$, the Schwinger result. The difference
between  the extremal  and the wormhole actions is $\pi \hq^2$ for
$a=0$ and zero for $0<a<1$ as $\hq\hB \rightarrow 0$.

\listrefs
\end